\documentstyle[preprint,eqsecnum,amsfonts,aps]{revtex}

\newcommand{\sous}[2]{\stackrel{\phantom{(n,p)}}{#2} \! \! \! \! \! \!
			\! \! \! 
			{{} \atop \scriptstyle #1} {}}

\newtheorem{theorem}{Theorem}
\newtheorem{lemma}{Lemma}
\newtheorem{proposition}{Proposition}
\newtheorem{definition}{Definition}

\begin{document}
\draft

\title{Hadamard Regularization}

\author{Luc Blanchet}
\address{D\'epartement d'Astrophysique Relativiste et de Cosmologie,\\
Centre National de la Recherche Scientifique (UMR 8629),\\
Observatoire de Paris, 92195 Meudon Cedex, France,\\
and Theoretical Astrophysics,
California Institute of Technology,\\
Pasadena, California 91125, U.S.A.} 

\author{Guillaume Faye}
\address{D\'epartement d'Astrophysique Relativiste et de Cosmologie,\\
Centre National de la Recherche Scientifique (UMR 8629),\\
Observatoire de Paris, 92195 Meudon Cedex, France}

\date{\today}
\maketitle
\begin{abstract}
Motivated by the problem of the dynamics of point-particles in high
post-Newtonian (e.g. 3PN) approximations of general relativity, we
consider a certain class of functions which are smooth except at some
isolated points around which they admit a power-like singular
expansion. We review the concepts of (i) Hadamard ``partie finie'' of
such functions at the location of singular points, (ii) the partie
finie of their divergent integral. We present and investigate
different expressions, useful in applications, for the latter partie
finie. To each singular function, we associate a partie-finie (${\rm
Pf}$) pseudo-function. The multiplication of pseudo-functions is
defined by the ordinary (pointwise) product. We construct a
delta-pseudo-function on the class of singular functions, which
reduces to the usual notion of Dirac distribution when applied on
smooth functions with compact support. We introduce and analyse a new
derivative operator acting on pseudo-functions, and generalizing, in
this context, the Schwartz distributional derivative. This operator is
uniquely defined up to an arbitrary numerical constant. Time
derivatives and partial derivatives with respect to the singular
points are also investigated. In the course of the paper, all the
formulas needed in the application to the physical problem are
derived.
\end{abstract}

\section{Introduction}

The Hadamard regularization \cite{Hadamard,Schwartz}, based on the
concept of finite part (``partie finie'') of a singular function or a
divergent integral, plays an important role in several branches of
Mathematical Physics (see \cite{EstrK85,EstrK89,Sellier,Jones96} for
reviews). Typically one deals with functions admitting some
non-integrable singularities on a discrete set of isolated points
located at finite distances from the origin. The regularization
consists of assigning {\it by definition} a value for the function at
the location of one of the singular points, and for the (generally
divergent) integral of that function. The definition may not be fully
deterministic, as the Hadamard partie finie depends in general on some
arbitrary constants. The Hadamard regularization is one among several
other possible regularizations \cite{EstrK89}.

A motivation for investigating the properties of a regularization
comes from the physical problem of the gravitational interaction of
compact bodies in general relativity. As it is hopeless to find a
sufficiently general exact solution of this problem, we resort to
successive post-Newtonian approximations (limit $c\to
+\infty$). Within the post-Newtonian framework, it makes sense to
model compact objects like black holes by point-like particles. This
is possible at the price of introducing a regularization, in order to
cure the divergencies due to the infinite self-field of the
point-masses. However, general relativity is a non-linear theory and,
if we want to go to high post-Newtonian approximations, involving high
non-linear terms, the process of regularization must be carefully
defined. In particular, it turns out that, from the
third-post-Newtonian approximation (3PN or $1/c^6$), the problem
becomes complicated enough that a rather sophisticated version of the
Hadamard regularization, including a theory of generalized functions,
is required. By contrast, a cruder form of the Hadamard
regularization, using merely the concept of partie finie of singular
functions \cite{BeDD81,DD81a,S85,BDI95,Jara97,JaraS98,BFP98}, is
sufficient to treat the problem up to the 2PN order.  Furthermore, we
know that the answer provided by the Hadamard regularization up to the
2PN order is correct, in the sense that the field of the two bodies
matches the inner field generated by two black holes \cite{D83a}, and
the result for the equations of motion can be recovered without need
of any regularization from computations valid for extended
non-singular objects \cite{Kop85,GKop86}.  Conforted by these
observations we systematically investigate in this paper the Hadamard
regularization as well as a theory of associated generalized
functions, in a form which can be directly applied to the study of the
dynamics of two point-like particles at the 3PN order \cite{BF00}. (We
therefore restrict our attention to two singular points; however most
of the results of the paper can be generalized to any number of
points.)  Notice that this problem enjoys a direct relevance to the
future gravitational-wave experiments LIGO and VIRGO, which should be
able to detect the radiation from black-hole and/or neutron-star
binaries which a precision compatible with the 3PN approximation
\cite{3mn}.

Consider the class ${\cal F}$ of functions on ${\mathbb R}^3$ that are
smooth except at two isolated singularities 1 and 2, around which they
admit some power-like singular expansions.  The Hadamard partie finie
$(F)_1$ of $F\in{\cal F}$ at the location of singularity 1, as
reviewed in Section II, is defined by the average over spatial
directions of the finite-part coefficient in the expansion of $F$
around 1. On the other hand, the Hadamard partie finie ${\rm Pf}\int
d^3{\bf x}~\!F$ of the divergent integral of $F$, we will review in
Section III, is obtained from the removal to the integral of the
divergent part arising when two regularizing volumes surrounding the
singularities shrink to zero. Both concepts of partie finie are
closely related. Notably, the partie-finie integral of a gradient is
equal to the sum of the parties finies (in the former sense) of the
surface integrals surrounding the singularities, in the limit of
vanishing areas. In Section IV we investivage several alternative
expressions of the Hadamard partie finie of integrals, some of them
based on a finite part defined by means of an analytic continuation
process (see \cite{Schwartz} for a relation between partie finie and
analytic continuation). In our terminology, we adopt the name ``partie
finie'' for the specific definitions due to Hadamard, and speak of a
``finite part'' when referring to other definitions, based for
instance on analytic continuation. In Section V we focus to the case
(important in applications) of the partie finie of a Poisson integral
of $F\in {\cal F}$.

To any $F\in {\cal F}$, we associate in Section VI a generalized
function, or partie-finie ``pseudo-function'' ${\rm Pf}F$, which is a
linear form on ${\cal F}$ defined for any $G\in{\cal F}$ by the
duality bracket $<{\rm Pf}F,G>={\rm Pf}\int d^3{\bf x}\!~FG$. When
restricted to the set ${\cal D}$ of smooth functions with compact
support the pseudo-function ${\rm Pf}F$ is a distribution in the sense
of Schwartz \cite{Schwartz} (see also \cite{Gelfand,Jones82,Kanwal83}
for more details about generalized functions and distributions),
i.e. a linear form which is continuous with respect to the Schwartz
topology. [However, we do not attempt here to introduce a topology on
${\cal F}$; we simply define the set of algebraic and differential
rules, needed in applications, that are satisfied by the
pseudo-functions on ${\cal F}$.] The product of pseudo-functions
coincides with the ordinary (``pointwise'') product used in Physics,
namely ${\rm Pf}F\!~.\!~{\rm Pf}G={\rm Pf}(FG)$. An important
particular case is the pseudo-function ${\rm Pf}\delta_1$ obtained (in
Section VI) from the pseudo-function associated with the Riesz
delta-function \cite{Riesz}, and that satisfies $\forall G\in{\cal
F}$, $<{\rm Pf}\delta_1,G>=(G)_1$. The ``Dirac pseudo-function'' ${\rm
Pf}\delta_1$ plays in the present context the same role as plays the
Dirac measure in distribution theory. We introduce also more
complicated objects such as ${\rm Pf}(F\delta_1)$. In Sections VII and
VIII we show how to construct a derivative operator on ${\cal F}$,
generalizing for this class of function the standard distributional
derivative operator on ${\cal D}$ and satisfying basically the
so-called rule of integration by parts, namely $\forall F,G\in {\cal
F}$, $<\partial_i({\rm Pf}F),G>=\!-\!<\partial_i({\rm Pf}G),F>$. In
addition we require that the derivative reduces to the ``ordinary''
derivative for functions that are bounded in a neighbourhood of the
singular points, and that the rule of commutation of derivatives
holds. We find that this derivative operator is uniquely defined
modulo a dependence on an arbitrary numerical constant (see Theorem 4
in Section VIII). It represents a natural notion of derivative within
the context of Hadamard regularization of the functions in ${\cal
F}$. However, it does not satisfy in general the Leibniz rule for the
derivative of a product (in agreement with a theorem of Schwartz
\cite{Schwartz54}). See Colombeau \cite{Colombeau} for a
multiplication of distributions and associated distributional
derivative satisfying the Leibniz rule. Further, we obtain the rules
obeyed by the new derivative operator when acting on pseudo-functions
such as ${\rm Pf}(F\delta_1)$ in Section VII, and we investigate the
associated Laplacian operator in Section VIII. Finally, in Section IX,
we consider the case of partial derivatives with respect to the
singular points 1 and 2, as well as the time derivative when both
singular points depend on time (i.e. represent the trajectories of
real particules). Within this approach, the latter distributional
derivative constitutes an important tool when studying the problem of
the gravitational dynamics of point-particles at the 3PN order
\cite{BF00}.

\bigskip\noindent
{\it Notation.} ${\mathbb N}$, ${\mathbb Z}$, ${\mathbb R}$ and
${\mathbb C}$ are the usual sets of non-negative integers, integers,
real numbers and complex numbers; ${\mathbb R}^{+*}$ is the set of
strictly positive real numbers $s>0$; ${\mathbb R}^3$ is the usual
three-dimensional space endowed with the Euclidean norm $|{\bf
x}|=(x_1^2+x_2^2+x_3^2)^{1/2}$; $C^p(\Omega)$ is the set of $p$-times
continuously differentiable functions on the open set $\Omega$ ($p\leq
+\infty$); $L^1_{\rm loc}(\Omega)$ is the set of locally integrable
functions on $\Omega$; the $o$ and $O$ symbols for remainders have
their standard meaning; distances between the field point ${\bf x}$
and the source points ${\bf y}_1$ and ${\bf y}_2$ are denoted by
$r_1=|{\bf x}-{\bf y}_1|$ and $r_2=|{\bf x}-{\bf y}_2|$; unit
directions are ${\bf n}_1=({\bf x}-{\bf y}_1)/r_1$ and ${\bf
n}_2=({\bf x}-{\bf y}_2)/r_2$; $d\Omega_1$ and $d\Omega_2$ are the
solid angle elements associated with ${\bf n}_1$ and ${\bf n}_2$;
$r_{12}=|{\bf y}_1-{\bf y}_2|$; ${\cal B}_1(s)$ and ${\cal B}_2(s)$
denote the closed spherical balls of radius $s$ centered on ${\bf
y}_1$ and ${\bf y}_2$; $\partial_i=\partial/\partial x^i$,
${}_1\partial_i=\partial/\partial y_1^i$,
${}_2\partial_i=\partial/\partial y_2^i$; $L=i_1i_2\cdots i_l$ is a
multi-index with length $l$; $n_1^L=n_1^{i_1}\cdots n_1^{i_l}$ and
$\partial_L=\partial_{i_1}\cdots \partial_{i_l}$; the
symmetric-trace-free (STF) projection is denoted by ${\hat n}_1^L={\rm
STF}(n_1^L)$; $(ij)=\case{ij+ji}{2}$ and $[ij]=\case{ij-ji}{2}$;
$1\leftrightarrow 2$ means the same expression but corresponding to
the point 2; for clearer reading, we use left-side labels 1 and 2 when
the quantity appears within the text, like for the partial derivatives
${}_1\partial_i$ and ${}_2\partial_i$ or the coefficients ${}_1f_a$
and ${}_2f_b$, and labels placed underneath the quantity when it
appears in an equation; iff means if and only if.

\section{Hadamard partie finie}

\subsection{A class of singular functions}

All over this paper we consider the class of functions of a ``field''
point ${\bf x}\in {\mathbb R}^3$ that are singular at the location of
two ``source'' points ${\bf y}_1$ and ${\bf y}_2$ around which they
admit some singular expansions.

\begin{definition}
A real function $F({\bf x})$ on ${\mathbb R}^3$ is said to belong to
the class of functions ${\cal F}$ iff

(i) $~F$ is smooth on ${\mathbb R}^3$ deprived from ${\bf y}_1$ and
${\bf y}_2$, i.e. $F\in C^\infty ({\mathbb R}^3-\{{\bf y}_1,{\bf
y}_2\})$.

(ii) There exists an ordered family of indices $(a_i)_{i\in {\mathbb
N}}$ with $a_i\in {\mathbb R}$, and a family of coefficients
$\!\!\sous{\!1}{~f}_{~a_i}$, such that

\begin{equation}\label{4}
\forall N\in {\mathbb N}\;,\qquad
F({\bf x})=\sum_{i=0}^{i_N} r_1^{a_i} \!\!\sous{1}{f}_{a_i}({\bf n}_1)
+ \!\!\sous{1}{R}_{~N}({\bf x}) \;.
\end{equation}
Here $r_1=|{\bf x}-{\bf y}_1|$ and ${\bf n}_1=({\bf x}-{\bf
y}_1)/r_1$; $i_N$ satisfies $a_0 < a_1 <\cdots < a_{i_N} \leq N <
a_{i_N+1}$; and the ``remainder'' is

\begin{equation}\label{5}
\sous{1}{R}_{~N}({\bf x}) = o(r_1^N)\qquad\mbox{when $~~r_1\to 0$} \;.
\end{equation} 

(iii) Idem with indices $(b_i)_{i\in {\mathbb N}}$, coefficients
$\!\!\sous{\!2}{~f}_{~b_i}$, remainder ${}_2R_N$, $r_1\leftrightarrow
r_2$ and ${\bf n}_1\leftrightarrow {\bf n}_2$.
\end{definition}
In addition to Definition 1, we always assume that the functions $F\in
{\cal F}$ decrease sufficiently fast at infinity (when $|{\bf
x}|\to\infty$) so that all integrals we meet are convergent at
infinity. Thus, when discussing the integral $\int d^3{\bf x}\!~F$, we
suppose implicitly that $F=o(|{\bf x}|^{-3})$ at infinity [or
sometimes $F=O(|{\bf x}|^{-3-\epsilon})$ where $\epsilon>0$], so that
the possible divergencies come only from the bounds at the singular
points ${\bf y}_{1,2}$. Similarly, when considering the integral $\int
d^3{\bf x}\!~FG$, we suppose $FG=o(|{\bf x}|^{-3})$, but for instance
we allow $F$ to blow up at infinity, say $F=O(|{\bf x}|)$, if we know
that $G$ decreases rapidly, e.g. $G=o(|{\bf x}|^{-4})$; in the case of
$\int d^3{\bf x}\!~\partial_iF$, we generally assume $F=o(|{\bf
x}|^{-2})$.  [Clearly, from Definition 1 the ordinary product $FG$ of
two functions of ${\cal F}$ is again a function of ${\cal F}$; and
similarly the ordinary gradient $\partial_iF\in {\cal F}$.]

An important assumption in Definition 1 is that the powers of $r_1$ in
the expansion of $F$ when $r_1\to 0$ (and similarly when $r_2\to 0$)
are bounded from below, i.e. $a_0\leq a_i$ where the most
``divergent'' power of $r_1$, which clearly depends on $F$, is
$a_0=a_0(F)$. Thus the part of the expansion which diverges when
$r_1\to 0$ is composed of a finite number of terms. Notice also that
we have excluded in Definition 1 the possible appearance of logarithms
of $r_1$ (or $r_2$) in the expansion of $F$. See Sellier
\cite{Sellier} for a more general study in the case where some
arbitrary powers of logarithms are present. We will discuss the
occurence of logarithms in Section V, when dealing with the Poisson
integral of $F$. At last, we point out that the coefficients ${}_1f_a$
(and similarly ${}_2f_b$) do not depend only on ${\bf n}_1$, but also
they do on the source points ${\bf y}_1$ and ${\bf y}_2$, so that in
principle we should write ${}_1f_a({\bf n}_1;{\bf y}_1,{\bf y}_2)$;
however, for simplicity's sake we omit writing the dependence on the
source points. The coefficients could also depend on other variables
such as the velocities ${\bf v}_1$ and ${\bf v}_2$ of the source
points, but the velocities do not participate to the process of
regularization and can be ignored for the moment (we will return to
this question in Section IX when considering the time dependence of
$F$).

Once the class ${\cal F}$ has been defined, we shall often write in
this paper the expansions of $F$ when $r_{1,2}\to 0$ in the simplified
forms

\begin{mathletters}\label{6}\begin{eqnarray}
F({\bf x})&=&\sum_{a_0\leq a\leq N} r_1^a \!\!\sous{1}{f}_a({\bf n}_1)
+ o(r_1^N)\qquad\mbox{when $r_1\to 0$} \;,\\ F({\bf
x})&=&\sum_{b_0\leq b\leq N} r_2^b \!\!\sous{2}{f}_b({\bf n}_2) +
o(r_2^N)\qquad~\mbox{when $r_2\to 0$} \;,
\end{eqnarray}\end{mathletters}$\!\!$
by which we really mean the expansions in Definition 1, i.e. in
particular where the indices $a\in (a_i)_{i\in {\mathbb N}}$ and $b\in
(b_i)_{i\in {\mathbb N}}$, and are {\it a priori} real. However, most
of the time (in applications), it is sufficient to assume that the
powers of $r_{1,2}$ are relative integers $a,b\in {\mathbb Z}$. We can
then write the expansion $r_1\to 0$ in the form

\begin{equation}\label{6'}
F=\sum_{k=0}^{k_0} {1\over r_1^{1+k}} \!\!\sous{1}{f}_{-1-k}+\sum_{k=0}^
N r_1^k \!\!\sous{1}{f}_k+ o(r_1^N)\;,
\end{equation}
where $k_0=-1-a_0$. In the following we shall sometimes derive the
results in the simpler case where the powers $\in {\mathbb Z}$, being
always undertood that the generalization to the case of real powers is
straightforward.  Finally, it is worth noting that the assumption (i)
in Definition 1, that $F$ is $C^\infty$ outside $\{{\bf y}_1,{\bf
y}_2\}$, can often be relaxed to allow some functions to have
integrable singularities. An example is the function ${\bf
x}\rightarrow 1/|{\bf x}-{\bf x}'|$ encountered in Section V,
depending on a fixed ``spectator'' point ${\bf x}'$ distinct from
${\bf y}_1$ and ${\bf y}_2$. To treat such objects, we introduce a
larger class of functions, ${\cal F}_{\rm loc}$.

\begin{definition}
$F({\bf x})$ is said to belong to the class of functions ${\cal
F}_{\rm loc}$ iff

(i') $~F$ is locally integrable on ${\mathbb R}^3$ deprived from ${\bf
y}_1$ and ${\bf y}_2$, i.e. $F\in L^1_{\rm loc}({\mathbb R}^3-\{{\bf
y}_1,{\bf y}_2\})$.

(ii)-(iii) in Definition 1 hold.
\end{definition}
For simplicity, in the following, we shall derive most of the results
for functions belonging to the class ${\cal F}$ (even if the
generalization to ${\cal F}_{\rm loc}$ is trivial); ${\cal F}_{\rm
loc}$ will be employed only occasionally.

\subsection{Partie finie of a singular function}

The first notion of Hadamard partie finie is that of a singular
function at the very location of one of its singular points.

\begin{definition}
Given $F\in {\cal F}$ we define the Hadamard partie finie of $F$ at
the point ${\bf y}_1$ to be
 
\begin{equation}\label{7}
(F)_1 = \int {d\Omega_1\over 4\pi} \sous{1}{f}_0({\bf n}_1)\;,
\end{equation}
where $d\Omega_1=d\Omega ({\bf n}_1)$ denotes the solid angle element
of origin ${\bf y}_1$ and direction ${\bf n}_1$.
\end{definition}
In words, the partie finie of $F$ at point 1 is defined by the angular
average, with respect to the unit direction ${\bf n}_1$, of the
coefficient of the zeroth power of $r_1$ in the expansion of $F$ near
1 (and similarly for the point 2). There is a non zero partie finie
only if the family of indices $(a_i)_{i\in {\mathbb N}}$ in Definition
1 contains the value 0, i.e. $\exists i_0$ such that $a_{i_0}=0$.  The
latter definition applied to the product $FG$ of two functions in
${\cal F}$ yields

\begin{equation}\label{8}
(FG)_1 = \sum_{a_0(F)\leq a\leq -a_0(G)} ~\int {d\Omega_1\over 4\pi}
\!\!\sous{1}{f}_a \!\!\sous{1}{g}_{-a}\;,
\end{equation}
where ${}_1f_a$ and ${}_1g_a$ are the coefficients in the expansions
of $F$ and $G$ when $r_1\to 0$ (the summation over $a$ is always
finite). From (\ref{8}) it is clear that the Hadamard partie finie is
not ``distributive'' with respect to the multiplication, in the sense
that

\begin{equation}\label{9}
(FG)_1 \not= (F)_1 (G)_1\quad\mbox{in general}\;.
\end{equation}

The partie finie picks up the angular average of ${}_1f_0({\bf n}_1)$,
namely the scalar or $l=0$ piece in the spherical-harmonics expansion
($Y_{lm}$), or, equivalently, in the expansion on the basis of
symmetric and trace-free (STF) products of unit vectors ${\bf
n}_1=(n_1^i)$. For any $l\in {\mathbb N}$, we denote by
$L=i_1i_2\cdots i_l$ a multi-index composed of $l$ indices, and
similarly $L-1=i_1i_2\cdots i_{l-1}$, $P=j_1j_2\cdots j_p$. In general
we do not need to specify the carrier index $i$ or $j$, so a tensor
with $l$ upper indices is denoted $T^L$, and for instance the scalar
formed by contraction with another tensor $U^L$ of the same type is
written as $S=T^L U^L=T^{i_1\cdots i_l}U^{i_1\cdots i_l}$, where we
omit writing the $l$ summations over the $l$ indices $i_k=1,2,3$. We
denote a product of $l$ components of the unit vector $n_1^i$ by
$n_1^L=n_1^{i_1}\cdots n_1^{i_l}$, and the STF projection of that
product by ${\hat n}_1^L\equiv {\rm STF}(n_1^L)$: e.g., ${\hat
n}_1^{ij}=n_1^i n_1^j-\case{1}{3}\delta^{ij}$, ${\hat n}_1^{ijk}=n_1^i
n_1^j n_1^k-\case{1}{5}(n_1^i \delta^{jk}+n_1^j \delta^{ki}+n_1^k
\delta^{ij}$). More generally, we denote by ${\hat T}^L$ the STF
projection of $T^L$; that is, ${\hat T}^L$ is symmetric, and satisfies
$\delta_{i_{l-1}i_l} {\hat T}^{i_{l-1}i_lL-2}=0$ (see \cite{Th80} and
the appendix A of \cite{BD86} for a compendium of formulas using the
STF formalism). The coefficients ${}_1f_a$ of the expansion of $F$
admit the STF decomposition

\begin{equation}\label{10}
\sous{1}{f}_a({\bf n}_1)=\sum_{l=0}^{+\infty} ~n_1^L\!\!\sous{1}{{\hat f}}_a^{~L}\;,
\end{equation}
where the ${}_1{\hat f}_a^L$'s are constant STF tensors, given by the
inverse formula:

\begin{equation}\label{11}
\sous{1}{{\hat f}}_a^{~L}={(2l+1)!!\over l!}\int{d\Omega_1\over 4\pi}
~{\hat n}_1^L \!\!\sous{1}{f}_a({\bf n}_1)\;.
\end{equation}
In STF notation, the Hadamard partie finie of $F$ at 1 reads simply

\begin{equation}\label{12}
(F)_1=\sous{1}{{\hat f}}_0\;,
\end{equation}
where ${}_1{\hat f}_a$ denotes the first term in the expansion (\ref{10}). 

\begin{lemma} 
The partie finie at 1 of the gradient $\partial_iF$ (as defined
outside the singularities) of any function $F\in {\cal F}$ satisfies
 
\begin{equation}\label{16}
(\partial_iF)_1=3\biggl({n_1^i\over r_1}F\biggr)_1\;.
\end{equation}
\end{lemma}
This Lemma is particularly useful as it permits replacing
systematically the differential operator $\partial_i$ by the {\it
algebraic} one $3\case{n_1^i}{r_1}$ when working under the
partie-finie sign $(..)_1$.

\bigskip\noindent 
{\it Proof.} The expansion when $r_1\to 0$ of the gradient is readily
obtained from the expansion of $F$ itself as
 
\begin{equation}\label{13}
\partial_iF = \sum_a r_1^{a-1} \biggl[a ~n_1^i \!\!\sous{1}{f}_{a} + 
d_1^{~\!\!i} \!\!\!\!\sous{1}{f}_{a}\biggr]\;,
\end{equation}
(with over-simplified notation for the sum), where the operator
$d_1^{~\!\!i}$ is defined as $r_1 \partial_i$ when applied on a
function of the sole unit vector ${\bf n}_1$. Hence, explicitly,
$d_1^{~\!\!i} = (\delta^{ij}-n_1^{ij})\case{\partial}{\partial
n_1^j}$. This operator is evidently transverse to ${\bf n}_1$ : $n_1^i
d_1^{~\!\!i} = 0$, and we get, from the decomposition (\ref{10}),

\begin{equation}\label{13'}
d_1^{~\!\!i}\!\!\!\sous{1}{f}_a=\sum_{l=0}^{+\infty}
~l\Bigl(n_1^{L-1}\!\!\sous{1}{{\hat
f}}_a^{~iL-1}-n_1^{iL}\!\!\sous{1}{{\hat f}}_a^{~L}\Bigr)\;.
\end{equation}
Thus, by averaging over angles,

\begin{equation}\label{14}
\int {d\Omega_1\over 4\pi} ~d_1^{~\!\!i} \!\!\sous{1}{f}_a
={2\over 3}\!\!\sous{1}{{\hat f}}_a^{~i}=2 \int {d\Omega_1\over 
4\pi} ~n_1^i \!\!\sous{1}{f}_a\;.
\end{equation}
We readily deduce that the partie finie of the gradient (\ref{13}) is
given by

\begin{equation}\label{15}
(\partial_iF)_1= 3 \int {d\Omega_1\over 4\pi} ~n_1^i \!\!\sous{1}{f}_1
= \sous{1}{{\hat f}}_1^{~i}\qquad\hbox{({\it QED}).}
\end{equation}

As an example of application of Lemma 1, we can write, using an
operation by parts,
$(r_1^3\partial_iF)_1=[\partial_i(r_1^3F)-\partial_i(r_1^3)F]_1=[3n_1^i
r_1^2F-\partial_i(r_1^3)F]_1$, from which it follows that

\begin{equation}\label{17}
(r_1^3\partial_iF)_1=0\;.
\end{equation}
Another consequence of Lemma 1, resulting from two operations by
parts, is $(r_1^2\Delta F)_1=[3n_1^i
r_1\partial_iF-\partial_i(r_1^2)\partial_iF]_1=(n_1^i
r_1\partial_iF)_1=[3F-\partial_i(n_1^ir_1)F]_1$ (where the Laplacian
$\Delta = \partial_i \partial_i$), hence the identity

\begin{equation}\label{18}
(r_1^2\Delta F)_1=0\;.
\end{equation}
By the same method we obtain also

\begin{equation}\label{19}
(\partial_{ij}F)_1=\biggl({15n_1^{ij}-3\delta^{ij}\over
r_1^2}F\biggr)_1=2~(\!\!\sous{1}{{\hat f}}_2^{~ij}+\delta^{ij}
\!\!\!\sous{1}{{\hat f}}_2)\;,
\end{equation}
the right-hand side of the last equality being expressed in terms of
the STF tensors parametrizing (\ref{10}).  Tracing out the previous
formula, we find

\begin{equation}\label{20}
(\Delta F)_1=\biggl({6\over r_1^2}F\biggr)_1=6\!\!\sous{1}{{\hat f}}_2
\;.
\end{equation}
Finally, let us quote the general formula for the partie finie of the
$l$th derivative $\partial_L F=\partial_{i_1}\cdots \partial_{i_l} F$:

\begin{equation}\label{21}
(\partial_LF)_1=l!\sum_{k=0}^{\bigl[\case{l}{2}\bigr]}~\delta^{(2K}
\!\!\!\sous{1}{{\hat f}}_l^{~L-2K)}\;.
\end{equation}
Here, $\bigl[\case{l}{2}\bigr]$ denotes the integer part of
$\case{l}{2}$, $\delta^{2K}$ is the product of Kronecker symbols
$\delta^{i_1i_2}\delta^{i_3i_4}\cdots \delta^{i_{2k-1}i_{2k}}$, and
${}_1{\hat f}_l^{L-2K}={}_1{\hat f}_l^{i_{2k+1}\cdots i_l}$; the
parenthesis around the indices denote the symmetrization. One may
define the ``regular'' part of the function $F$ near the singularity 1
as the formal Taylor expansion when $r_1\to 0$ obtained using
(\ref{21}). Thus,

\begin{equation}\label{21'}
F_1^{\rm reg}\equiv\sum_{l=0}^{+\infty}{1\over l!}r_1^l
n_1^L(\partial_LF)_1=\sum_{l=0}^{+\infty}r_1^l\sum_{k=0}^{
\bigl[\case{l}{2}\bigr]}~n_1^{L-2K}
\!\!\!\sous{1}{{\hat f}}_l^{~L-2K}\;.
\end{equation}

\section{Partie-finie integrals}

\subsection{The partie finie of a divergent integral}

The second notion of Hadamard partie finie is that of the integral
$\int d^3{\bf x}\!~F({\bf x})$, where $F\in{\cal F}$. This integral is
generally divergent because of the presence of the singular points
${\bf y}_1$ and ${\bf y}_2$ (recall that we always assume that the
function decreases sufficiently rapidly at infinity so that we never
have any divergency coming from the integration bound $|{\bf x}|\to
+\infty$). Consider first the domain ${\mathbb R}^3$ deprived from two
spherical balls ${\cal B}_1(s)$ and ${\cal B}_2(s)$ of radius $s$,
centered on the two singularities ${\bf y}_1$, ${\bf y}_2$: ${\cal
B}_1(s)=\{ {\bf x}; ~r_1\leq s \}$ and ${\cal B}_2(s)=\{ {\bf x};
~r_2\leq s \}$. We assume that $s$ is small enough,
i.e. $s<\case{r_{12}}{2}$ where $r_{12}=|{\bf y}_1-{\bf y}_2|$, so
that the two balls do not intersect. For $s>0$ the integral over this
domain, say $I(s)=\int_{{\mathbb R}^3\setminus {\cal B}_1(s) \cup
{\cal B}_2(s)} d^3{\bf x}~F$, is well-defined and generally tends to
infinity when $s\to 0$. Thanks to the expansions (assumed in
Definition 1) of $F$ near the singularities, we easily compute the
part of $I(s)$ that blows up when $s\to 0$; we find that this
divergent part is given, near each singularity, by a finite sum of
strictly negative powers of $s$ (a polynomial of $1/s$ in general)
plus a term involving the logarithm of $s$. By subtracting from $I(s)$
the corresponding divergent part, we get a term that possesses a
finite limit when $s\to 0$; the Hadamard partie finie \cite{Hadamard}
is defined as this limit. Associated with the logarithm of $s$, there
arises an ambiguity which can be viewed as the freedom in the
re-definition of the unit system we employ to measure the length
$s$. In fact it is convenient to introduce two constant length scales
$s_1$ and $s_2$, one per singularity, in order to a-dimensionalize the
logarithms as $\ln (\case{s}{s_1})$ and $\ln (\case{s}{s_2})$.

\begin{definition}
For any $F\in {\cal F}$ integrable in a neighbourhood of $|{\bf
x}|=+\infty$, we define the Hadamard partie finie of the divergent
integral $\int d^3{\bf x}~F$ as

\begin{eqnarray}\label{21''}
{\rm Pf}_{s_1,s_2}\int d^3{\bf x}~F=\lim_{s\to
0}~\biggl\{\int_{{\mathbb R}^3\setminus {\cal B}_1(s)\cup {\cal
B}_2(s)} d^3{\bf x}~F&+&\sum_{a+3<0}{s^{a+3}\over a+3}\int d\Omega_1
\!\!\!\sous{1}{f}_a +\ln\left({s\over s_1}\right)\int d\Omega_1
\!\!\!\sous{1}{f}_{-3}\nonumber\\ &+&1\leftrightarrow 2\}\;,
\end{eqnarray}
where $1\leftrightarrow 2$ means the same previous two terms but concerning the
singularity 2. 
\end{definition}
This notion of partie finie can be extended to functions which are
locally integrable outside the singularities, i.e. $F\in {\cal F}_{\rm
loc}$ (see Definition 2).  In (\ref{21''}) the divergent terms are
composed of a sum over $a$ such that $a+3<0$ as well as a logarithmic
term, by which we really mean, using the more detailed notation of
Definition 1, $$
\sum_{i=0}^{i_l-1}{s^{a_i+3}\over a_i+3}\int d\Omega_1 \!\!\!\sous{1}{f}_{a_i}
+\delta_{-3,a_{i_l}}\ln\left({s\over s_1}\right)\int d\Omega_1 \!\!\!
\sous{1}{f}_{a_{i_l}}
+1\leftrightarrow 2\;, $$ where $i_l$ is such that $a_0<a_1<\cdots
<a_{i_l-1}<-3\leq a_{i_l}$ (the sum is always finite); we have
introduced a Kronecker symbol $\delta_{-3,a_{i_l}}$ to recall that the
logarithm is present only if the family of indices $(a_i)_{i\in
{\mathbb N}}$ contains the integer $-3$ (i.e. $a_{i_l}=-3$). The
divergent terms in (\ref{21''}) can also be expressed by means of the
partie finie defined by (\ref{7}). Indeed, they read $$
4\pi\biggl[~\!\!\sum_{a+3<0}{s^{a+3}\over a+3} \left({F\over
r_1^a}\right)_1 +\ln\left({s\over s_1}\right)
\left(r_1^3F\right)_1\biggr] +1\leftrightarrow 2 $$ [coming back to
the less detailed notation of (\ref{21''})].

The partie-finie integral (\ref{21''}) depends intrinsically on the
two arbitrary constants $s_1$ and $s_2$ introduced above. There is
another way to interpret these constants besides the necessity to take
into account the dimension of $s$, which is discussed by Sellier in
\cite{Sellier}. With this point of view we initially define the partie
finie using two arbitrarily shaped volumes ${\cal V}_1$ and ${\cal
V}_2$ instead of the two spherical balls ${\cal B}_1$ and ${\cal
B}_2$. Consider for instance the two volumes ${\cal V}_1=\{ {\bf x};
~r_1\leq s \rho_1({\bf n}_1) \}$ and ${\cal V}_2=\{ {\bf x}; ~r_2\leq
s \rho_2({\bf n}_2) \}$, where $s\in {\mathbb R}^{+*}$ measures the
size of the volumes and the two functions $\rho_1$ and $\rho_2$
describe their shape (the balls ${\cal B}_1$ and ${\cal B}_2$
corresponding simply to $\rho_1$ and $\rho_2\equiv 1$). Here, we
assume for simplicity that the volumes remain isometric to themselves
when $s$ varies. Then, the partie finie is defined as the limit of the
integral over ${\mathbb R}^3\setminus {\cal V}_1\cup {\cal V}_2$ to
which we subtract the corresponding divergent terms when $s\to 0$,
{\it without} adding any normalizing constant to the logarithms. In
this way, we find that the alternative definition is equivalent to our
definition (\ref{21''}) provided that $s_1$ and
$s_2$ are related to the {\it shapes} of the regularizing volumes
${\cal V}_1$ and ${\cal V}_2$ through the formula

\begin{equation}\label{22}
\ln s_1 \int d\Omega_1\!\!\sous{1}{f}_{-3}=\int d\Omega_1\!\!
\sous{1}{f}_{-3} \ln\rho_1\;,
\end{equation}
(and similarly for $s_2$). The arbitrariness on the two original
regularizing volumes is therefore encoded into the two (and only two)
constants $s_1$ and $s_2$. A closely related way to interpret them is
linked to the necessity to allow the change of the integration
variable ${\bf x}$ in the integral $\int d^3{\bf x}\!~F$. Such an
operation modifies the size and shape of the regularizing volumes,
thus the balls ${\cal B}_1$ and ${\cal B}_2$ are in general
transformed into some new volumes ${\cal V}_1$ and ${\cal V}_2$; so,
according to the previous argument, the freedom of choosing the
integration variable reflects out in the freedom of choosing two
arbitrary constants $s_1$ and $s_2$. (In this paper we shall assume
that $s_1$ and $s_2$ are fixed once and for all.)

An alternative expression of the Hadamard partie finie is often useful
because it does not involve the limit $s\to 0$, but is written with
the help of a {\it finite} parameter $s'\in {\mathbb
R}^{+*}$. Consider some $s'$ such that $0<s<s'$, and next, split the
integral over ${\mathbb R}^3\setminus {\cal B}_1(s)\cup {\cal B}_2(s)$
into the sum of the integral over ${\mathbb R}^3\setminus {\cal
B}_1(s')\cup {\cal B}_2(s')$ and the two integrals over the
ring-shaped domains ${\cal B}_1(s')\setminus {\cal B}_1(s)$ and ${\cal
B}_1(s')\setminus {\cal B}_1(s)$. If $s<s'\ll 1$ we can substitute
respectively into the ring-shaped integrals the expansions of $F$ when
$r_1\to 0$ and $r_2\to 0$ [see (\ref{6})]. The terms that are
divergent in $s$ cancel out, so we can apply the limit $s\to 0$ (with
fixed $s'$). This yields the following expression for the partie
finie: $\forall N\in {\mathbb N}$,

\begin{eqnarray}\label{22'}
{\rm Pf}_{s_1,s_2}\int d^3{\bf x}~F=\int_{{\mathbb R}^3\setminus {\cal
B}_1(s')\cup {\cal B}_2(s')} d^3{\bf x}~F&+&\sum_{a+3\leq N\atop
a+3\not= 0}{{s'}^{a+3}\over a+3}\int d\Omega_1 \!\!\!\sous{1}{f}_a
+\ln\left({s'\over s_1}\right)\int d\Omega_1
\!\!\!\sous{1}{f}_{-3}\nonumber\\ &+&1\leftrightarrow
2~+~o({s'}^{N+3})\;,
\end{eqnarray}
which is valid for an arbitrary fixed $s'$. Of course, up to any given
finite order $N$ the second member of (\ref{22'}) depends on $s'$, but
in the formal limit $N\to +\infty$, this dependence disappears and,
{\it in fine}, the partie finie is independent of $s'$.

\subsection{Partie-finie integral of a gradient}

A fundamental feature of the Hadamard partie finie of a divergent
integral is that the integral of a gradient $\partial_i F$ is {\it a
priori} not zero, since the surface integrals surrounding the two
singularities become infinite when the surface areas shrink to zero,
and may possess a finite part.

\begin{theorem}
For any $F\in {\cal F}$ the partie finie of the gradient of $F$ is
given by

\begin{equation}\label{23}
{\rm Pf}\int d^3{\bf x}~ \partial_iF=-4\pi ~(n_1^i
r_1^2F)_1+1\leftrightarrow 2\;,
\end{equation}
where the singular value at point 1 is defined by ${\rm (\ref{7})}$.
\end{theorem}
In the case of a regular function, the result is always zero from the
simple fact that the surface areas tend to zero --- {\it cf} the
factor $r_1^2$ in the right side of (\ref{23}). However, for $F\in
{\cal F}$, the factor $r_1^2$ is in general compensated by a divergent
term in the expansion of $F$, possibly producing a finite
contribution.

\bigskip\noindent 
{\it Proof.} We apply (\ref{21''}) to the case of the gradient
$\partial_iF$, using the expansion of $\partial_iF$ when $r_1\to 0$ as
given by (\ref{13}).  The expression of the divergent terms is
simplified with the help of the identity (\ref{14}), which shows
notably that the logarithms and associated constants $s_{1,2}$
disappear.  This leads to

\begin{equation}\label{26}
\lim_{s\to 0}~\biggl\{\int_{{\mathbb R}^3\setminus {\cal B}_1(s)\cup {\cal B}_2(s)}
d^3{\bf x}~\partial_iF+\sum_{a+2<0} s^{a+2}\int d\Omega_1 ~n_1^i
\!\!\sous{1}{f}_{a}+1\leftrightarrow 2\biggr\}\;.
\end{equation}
Next, the first term inside the braces is transformed via the Gauss
theorem into two surface integrals at $r_1=s$ and $r_2=s$, where we
can replace $F$ by the corresponding expansions around ${\bf y}_1$ and
${\bf y}_2$ respectively. We get

\begin{eqnarray*}
\lim_{s\to 0}~\biggl\{
-\sum_a s^{a+2}\int d\Omega_1 ~n_1^i \!\!\sous{1}{f}_{a}+
\sum_{a+2<0} s^{a+2}\int d\Omega_1 ~n_1^i \!\!\sous{1}{f}_{a}\biggr\}
&=&-\int d\Omega_1 ~n_1^i \!\!\sous{1}{f}_{-2}
\end{eqnarray*}
(and similarly when $1\leftrightarrow 2$); {\it QED}.

From Theorem 1 it results that the correct formula for ``integrating
by parts'' under the sign ${\rm Pf}$ is
  
\begin{equation}\label{24}
{\rm Pf}\int d^3{\bf x}~F\partial_iG=-{\rm Pf}\int d^3{\bf
x}~G\partial_iF-4\pi ~(n_1^i r_1^2FG)_1-4\pi ~(n_2^i r_2^2FG)_2\;.
\end{equation}
Note also that the partie-finie integrals of a double derivative as
well as a Laplacian are given by

\begin{mathletters}\label{25}\begin{eqnarray}
{\rm Pf}\int d^3{\bf x}~ \partial_{ij} F&=&4\pi ~\Bigl(r_1
(\delta^{ij}-2 n_1^{ij}) F\Bigr)_1+1\leftrightarrow 2\;,\\ {\rm
Pf}\int d^3{\bf x}~ \Delta F&=&4\pi ~(r_1 F)_1+1\leftrightarrow 2\;.
\end{eqnarray}\end{mathletters}$\!\!$

\subsection{Parties finies and the Riesz delta-function}

The Riesz delta-function \cite{Riesz} plays an important role in the
context of Hadamard parties finies. It is defined for any
$\varepsilon\in {\mathbb R}^{+*}$ by ${}_\varepsilon\delta ({\bf x}) =
\case{\varepsilon(1-\varepsilon)}{4\pi} |{\bf x}|^{\varepsilon-3}$;
when $\varepsilon\to 0$, it tends, in the usual sense of distribution
theory, towards the Dirac measure in three dimensions ---
i.e. $\lim_{\varepsilon\to 0} {}_\varepsilon\delta=\delta$, as can be
seen from the easily checked property that $\Delta (|{\bf
x}|^{\varepsilon-1})=-4\pi ~{}_\varepsilon\delta ({\bf x})$.  The
point for our purpose is that when defined with respect to one of the
singularities, the Riesz delta-function belongs to ${\cal F}$. Thus,
let us set, $\forall\varepsilon\in {\mathbb R}^{+*}$,

\begin{equation}\label{27}
{}_\varepsilon\delta_1({\bf x}) \equiv {}_\varepsilon\delta ({\bf
x}-{\bf y}_1)= {\varepsilon(1-\varepsilon)\over 4\pi}
r_1^{\varepsilon-3} ~\in~{\cal F}
\end{equation}
(and {\it idem} for 2). Now we can apply to
${}_\varepsilon\delta_1({\bf x})$ the previous definitions for parties
finies. In particular, from Definition 3, we see that
${}_\varepsilon\delta_1$ has no partie finie at 1 when $\varepsilon$
is small enough: $\bigl({}_\varepsilon\delta_1\bigr)_1=0$. From
Definition 4:

\begin{lemma}
For any $F\in {\cal F}$, we have

\begin{equation}\label{28}
\lim_{\varepsilon\to 0}~{\rm Pf}\int d^3{\bf x}~ 
{}_\varepsilon\delta_1 F = (F)_1 \;,
\end{equation}
where the value of $F$ at point 1 is given by the prescription ${\rm (\ref{7})}$. 
\end{lemma}

\noindent
{\it Proof.}  For $\varepsilon>0$ we evaluate the finite part of the
integral for the product ${}_\varepsilon\delta_1 F \in {\cal F}$ using
the specific form (\ref{22'}) of the partie finie defined in terms of
a given finite $s'$. The expansions of ${}_\varepsilon\delta_1 F$ when
$r_{1,2}$ tend to zero are readily determined to be

\begin{mathletters}\label{29}\begin{eqnarray} 
{}_\varepsilon\delta_1 F &=& \frac{\varepsilon (1-\varepsilon)}{4 \pi}
\sum_a r_1^{a+\varepsilon-3} \! \! \!
\! \sous{1}{f}_a({\bf n}_1) \qquad \qquad \qquad \qquad
\qquad \quad \; \; \, \hbox{for $r_1 \to 0$} 
\; , \\
{}_\varepsilon\delta_1 F &=&
\frac{\varepsilon (1-\varepsilon)}{4 \pi} \sum_{l \ge 0} \frac{(-)^l}{l!}
\!\!\sous{1}{\partial}_{~L} r_{12}^{\varepsilon-3} \sum_b
r_2^{b+l} n_2^L \! \! \! \! \sous{2}{f}_b({\bf n}_2) \qquad\quad\qquad
\hbox{for $r_2 \to 0$} \;.
\end{eqnarray}\end{mathletters}$\!\!$
In the second equation we used the Taylor expansion
$r_1^{\varepsilon-3}=\sum_{l\geq 0}\case{(-)^l}{l!}r_2^ln_2^L
{}_1\partial_L r_{12}^{\varepsilon-3}$ when $r_2\to 0$, with notation
$n_2^L=n_2^{i_1}\cdots n_2^{i_l}$ and
${}_1\partial_L={}_1\partial_{i_1}\cdots{}_1\partial_{i_l}$.  Hence,
we can write the partie-finie integral in the form ($\forall N\in
{\mathbb N}$; with fixed $s'$ such that $0<s'<1$)

\begin{eqnarray*} 
\int_{{\mathbb R}^3\setminus {\cal B}_1(s')\cup {\cal B}_2(s')}
d^3{\bf x}~{}_\varepsilon\delta_1 F&+&{\varepsilon
(1-\varepsilon)\over 4\pi}
\sum_{a+\varepsilon\leq N} \frac{{s'}^{a+\varepsilon}}{a+\varepsilon}
\int d\Omega_1 \! \! \! \! \sous{1}{f}_a\\ 
&+&  {\varepsilon (1-\varepsilon)\over 4\pi} \sum_{l \ge 0} \frac{(-)^l}{l!}
\!\!\sous{1}{\partial}_{~L} r_{12}^{\varepsilon-3} \biggl[ \sum_{b+l+3\leq 
N\atop {\rm and}\not=0} 
 \frac{{s'}^{b+l+3}}{b+l+3}
\int d\Omega_2 ~n_2^L \! \! \! \!
\sous{2}{f}_b \\
&&\qquad\qquad\qquad\qquad\qquad+\ln\biggl({s'\over s_2}\biggr)\int d\Omega_2
~n_2^L \! \! \! \! \sous{2}{f}_{-l-3} \biggr]+o({s'}^N) \; .  
\end{eqnarray*}
Here, we have discarded the term with $\ln\left(\case{s'}{s_1}\right)$
by choosing $\varepsilon>0$ to be so small that all denominators
$a+\varepsilon$ differ from zero. Since ${}_\varepsilon\delta_1$ tends
towards the Dirac measure when $\varepsilon \to 0$, the integral over
${{\mathbb R}^3\setminus {\cal B}_1(s')\cup {\cal B}_2(s')}$ goes to
zero. Because of the factor $\varepsilon$ present in the numerators,
so do the other terms when $\varepsilon \to 0$, except for those whose
denominators involve a compensating $\varepsilon$. Now, the only term
having the required property corresponds to $a=0$ in the previous
expression.  Therefore, taking the limit $\varepsilon\to 0$ (with
fixed $s'$), we get $$
\lim_{\varepsilon \to 0} ~{\rm Pf}\int d^3{\bf x}~{}_\varepsilon\delta_1 F 
 =\int \frac{d\Omega_1}{4 \pi}
\! \! \! \! \sous{1}{f}_0({\bf n}_1)+o({s'}^N)\;,
$$ and this being true for any $N$, we conclude $$
\lim_{\varepsilon \to 0} ~{\rm Pf}\int d^3{\bf x}~{}_\varepsilon\delta_1 F 
 = \int \frac{d\Omega_1}{4 \pi}
\! \! \! \! \sous{1}{f}_0({\bf n}_1)= (F)_1\qquad\hbox{({\it QED}).}
$$

As we can infer from Lemma 2, the Riesz delta-function
${}_\varepsilon\delta_1$ should constitute in the limit
$\varepsilon\to 0$ an appropriate extension of the notion of Dirac
distribution to the framework of parties finies of singular functions
in ${\cal F}$. The precise definition of a ``partie-finie Dirac
function'' necessitates the introduction of the space of linear forms
on ${\cal F}$ and will be investigated in Section VI (see Definition
7).

\section{Alternative forms of the partie finie}

\subsection{Partie finie based on analytic continuation}

Practically speaking, the Hadamard partie-finie integral in the form
given by (\ref{21''}) is rather difficult to evaluate, because it
involves an integration over the complicated volume ${\mathbb
R}^3\setminus {\cal B}_1(s)\cup {\cal B}_2(s)$. Fortunately, there
exist several alternative expressions of the Hadamard partie finie,
which are much better suited for practical computations. The first one
is based on a double analytic continuation, with two complex
parameters $\alpha$, $\beta\in {\mathbb C}$, of the integral

\begin{equation}\label{30}
I_{\alpha,\beta} = \int d^3{\bf x}~\left(\frac{r_1}{s_1}\right)^\alpha
\left(\frac{r_2}{s_2}\right)^\beta F\;,
\end{equation}
where the constants $s_1$ and $s_2$ are the same as those introduced
within the definition (\ref{21''}).  The point for our purpose is that
the integral (\ref{30}) does range over the complete set ${\mathbb
R}^3$. First of all, we propose to check that $I_{\alpha,\beta}$ is
defined by analytic continuation in a neighbourhood of the origin
$\alpha=0=\beta$ in ${\mathbb C}^2$, except at the origin itself where
it generically admits a simple pole in $\alpha$ or $\beta$ or both.
We start by splitting $I_{\alpha,\beta}$ into three contribution:
${}_1I_{\alpha,\beta}$ extending over the ball ${\cal B}_1(s)$ of
radius $s$ surrounding 1, ${}_2I_{\alpha,\beta}$ extending over the
ball ${\cal B}_2(s)$ surrounding 2, and ${}_3I_{\alpha,\beta}$
extending over the rest ${\mathbb R}^3\setminus {\cal B}_1(s)\cup
{\cal B}_2(s)$. The integral ${}_1I_{\alpha,\beta}$ is initially
convergent for $\Re (\alpha)>-a_0-3$ and any $\beta$, where $a_0$ is
the most singular power of $r_1$ in the expansion of $F$ near ${\bf
y}_1$; similarly, ${}_2I_{\alpha,\beta}$ exists only if $\Re
(\beta)>-b_0-3$ and any $\alpha$ ($b_0$ is the analogous to $a_0$ that
relates to ${\bf y}_2$), and ${}_3I_{\alpha,\beta}$ exists if $\Re
(\alpha+\beta)<\epsilon$, where $\epsilon>0$ is such that $F=O(|{\bf
x}|^{-3-\epsilon})$ when $|{\bf x}|\to +\infty$.  As the third
contribution ${}_3I_{\alpha,\beta}$ is clearly defined in a
neighbourhood of the origin, including the origin itself, we consider
simply the part ${}_1I_{\alpha,\beta}$ (the same reasoning applies to
${}_2I_{\alpha,\beta}$). Within the integrand, we replace the product
$r_2^\beta F$ by its expansion in the neighbourhood of ${\bf y}_1$
(using a Taylor expansion for $r_2^\beta$), and find that the
dependence on $\beta$ occurs through some everywhere well-defined
quantity, namely ${}_1\partial_Lr_{12}^\beta$. After performing the
angular integration over $d\Omega_1$, we obtain a remaining radial
integral consisting of a sum of terms of the type $\int_0^s dr_1
r_1^{\alpha+a+l+2}=s^{\alpha+a+l+3}/(\alpha+a+l+3)$, that clearly
admit a unique analytic continuation on ${\mathbb C}\!\setminus
\!{\mathbb Z}$; hence our statement (a simple pole at the origin
arises when $a=-l-3$).

\begin{theorem} 
For any function $F\in {\cal F}$ that is summable at infinity, the
Hadamard partie finie of the integral is given by

\begin{equation}\label{31}
{\rm Pf}_{s_1,s_2} \int d^3{\bf x}~F= {\rm FP}_{\alpha \to 0 \atop
\beta \to 0} \int d^3{\bf x}~
\left(\frac{r_1}{s_1} \right)^\alpha \left(\frac{r_2}{s_2} \right)^\beta F =
{\rm FP}_{\beta \to 0 \atop \alpha \to 0} \int d^3{\bf x}~
\left(\frac{r_1}{s_1} \right)^\alpha \left(\frac{r_2}{s_2}
\right)^\beta F \; ,
\end{equation}
where ${\rm FP}_{\alpha \to 0 \atop \beta \to 0}$ means taking the
finite parts in the Laurent expansions when $\alpha\to 0$ and
$\beta\to 0$ successively.
\end{theorem}
The proof of Theorem 2 is relegated to Appendix A. Notice our
convention regarding the notation: while ``${\rm Pf}$'' always stands
for the Partie finie of an integral in the specific sense of Hadamard
\cite{Hadamard}, we refer to ``${\rm FP}$'' as the Finite Part or
zeroth-order coefficient in the Laurent expansion with respect to some
complex parameter ($\alpha$, $\beta\in {\mathbb C}$, or $B\in {\mathbb
C}$ as in the next subsection). We see from Theorem 2 that the partie
finie ${\rm Pf}$ can be viewed as a finite part ${\rm FP}$ and {\it
vice versa}. The link between analytic continuation and Hadamard
partie finie is pointed out by Schwartz \cite{Schwartz}.  More
precisely, Theorem 2 says how to calculate the Hadamard partie finie;
the procedure consists of: (i) performing the Laurent expansion of
$I_{\alpha,\beta}$ when $\alpha\to 0$ while $\beta$ remains a {\it
fixed} (``spectator'') non-zero complex number, i.e.  $$
I_{\alpha,\beta}=\sum_{p=p_{\rm min}}^{+\infty}\alpha^p
\!~I_{(p),\beta}\;, $$ where $p\in \!{\mathbb Z}$ and where the
coefficients $I_{(p),\beta}$ depend on $\beta$; (ii) achieving the
Laurent expansion of the zeroth-$\alpha$-power coefficient
$I_{(0),\beta}$ when $\beta\to 0$, i.e.  $$
I_{(0),\beta}=\sum_{q=q_{\rm min}}^{+\infty}\beta^q \!~I_{(0,q)}\;, $$
to finally arrive at the zeroth-$\beta$-power coefficient
$I_{(0,0)}$. Indeed, we find that the same result can be obtained by
proceeding the other way around, first expanding around $\beta =0$
with a fixed $\alpha$, then expanding the coefficients
$I_{\alpha,(0)}$ near $\alpha =0$. Thus,

\begin{equation}\label{32} 
\sous{\!\!\!\!\beta\to 0}{\rm FP}\biggl\{\sous{\!\!\!\!\alpha\to 0}
{\rm FP}I_{\alpha,\beta} \biggr\} = I_{(0,0)}=\sous{\!\!\!\!\alpha\to 0}
{\rm FP}\biggl\{\sous{\!\!\!\!\beta\to 0}{\rm FP}I_{\alpha,\beta}\biggr\}\; .
\end{equation}
We emphasize that the definition (\ref{21''}) of the partie finie
yields unambiguously the result $I_{(0,0)}$, which corresponds to
taking {\it independently} the two limits $\alpha\to 0$ and $\beta\to
0$ (the limiting process does not allow for instance to keep
$\alpha=\beta$). The final value $I_{(0,0)}$ is the same as the one
given by the regularization adopted by Jaranowski and Sch\"afer
\cite{JaraS98} (see their appendix B.2).

In practice the expression (\ref{31}) is used in connection with the
Riesz formula \cite{Riesz}, valid for any $\gamma$, $\delta\in
{\mathbb C}$ except at some isolated poles,

\begin{equation}\label{33}
\int d^3{\bf x}~r_1^\gamma r_2^\delta=\pi^{3/2}{
\Gamma\left({\gamma+3\over 2}\right)\Gamma\left({\delta+3\over 2}\right)
\Gamma\left(-{\gamma+\delta+3\over 2}\right)\over
\Gamma\left(-{\gamma\over 2}\right)\Gamma\left(-{\delta\over 2}\right)
\Gamma\left({\gamma+\delta+6\over 2}\right)} r_{12}^{\gamma+\delta+3}\;,
\end{equation}
with $r_{12}=|{\bf y}_1-{\bf y}_2|$; here, $\Gamma$ denotes the
Eulerian function. According to Theorem 2, the formula (\ref{33})
permits computing the partie finie of any integral of a product
between powers of $r_1$ and $r_2$.  Consider the (not so trivial) case
of the integral of $r_1^{-3}r_2^{-3}$, which is divergent at both
points 1 and 2. From the Riesz formula, with $\gamma=\alpha-3$ and
$\delta=\beta-3$, we have $$ I_{\alpha,\beta}=\pi^{3/2}
{\Gamma\left({\alpha\over 2}\right)\Gamma\left({\beta\over 2}\right)
\Gamma\left(-{\alpha+\beta-3\over 2}\right)\over
\Gamma\left(-{\alpha-3\over 2}\right)\Gamma\left(-{\beta-3\over 2}\right)
\Gamma\left({\alpha+\beta\over 2}\right)} {r_{12}^{\alpha+\beta-3}\over 
s_1^\alpha s_2^\beta}\;.
$$ We compute the Laurent expansion when $\alpha\to 0$ with fixed
$\beta\in {\mathbb C}$ and obtain a simple pole in $\alpha$ followed
by a $\beta$-dependent finite part given by $$ I_{(0),\beta}=\pi^{3/2}
{\Gamma(1)\over\Gamma\left({3\over 2}\right)}{r_{12}^{\beta-3}\over
s_2^\beta}\left[ {2\over
\beta}+\Psi(1)-\Psi\left(1+\case{\beta}{2}\right)
+\Psi\left(\case{3}{2}\right)-\Psi\left(\case{3}{2}-\case{\beta}{2}\right)+2\ln
\left({r_{12}\over s_1}\right)\right]\;, $$ with $\Psi (z) =
\case{d}{dz}\ln\Gamma(z)$. This finite part itself includes a simple
pole in $\beta$, and then we obtain the corresponding finite part when
$\beta\to 0$ as $$ I_{(0,0)}={\pi^{3/2}\over
r_{12}^3}{\Gamma(1)\over\Gamma\left({3\over 2}\right)} \Biggl[ 2\ln
\left({r_{12}\over s_1}\right)+2\ln \left({r_{12}\over
s_2}\right)\Biggr]\;.  $$ At last, Theorem 2 tells us that

\begin{equation}\label{37}
{\rm Pf}_{s_1,s_2}\int {d^3{\bf x}\over r_1^3~\!r_2^3}={4\pi\over
r_{12}^3}\Biggl[\ln\left({r_{12}\over
s_1}\right)+\ln\left({r_{12}\over s_2}\right)\Biggr]\;.
\end{equation}
Some more complicated integrals will be obtained in the next subsection. 

\subsection{Partie finie based on angular integration}

The idea is to compute the partie-finie integral by performing an
angular integration, followed by the integration over some radial
variable. In a first stage, consider an integral that diverges at the
point 1, but converges at the point 2. According to (\ref{21''}), we
need to compute it over the domain ${\mathbb R}^3\setminus {\cal
B}_1(s)$; so it is natural to change the integration variable ${\bf
x}$ to ${\bf r}_1\equiv {\bf x}-{\bf y}_1$, carry on the angular
integration over $d\Omega_1=d\Omega ({\bf n}_1)$, and then, the radial
integration over $r_1=|{\bf r}_1|$ varying from $s$ to infinity, i.e.

\begin{equation}\label{38}
\int_{{\mathbb R}^3\setminus {\cal B}_1(s)}
d^3{\bf x}~F = \int_{r_1>s}
d^3{\bf r}_1~F = \int_s^{+\infty} dr_1 r_1^2 \int d\Omega_1~F \;.
\end{equation}
In the more general case where the integral is simultaneously
divergent at the two points 1 and 2, this method {\it stricto sensu}
is no longer valid since the radial integration in (\ref{38}) becomes
divergent when $r_1=r_{12}$. Yet, still it is advantageous to dispose
of a mean to change the variable ${\bf x}$ into ${\bf r}_1$ in order
to obtain a convenient radial integration (even at the price of
breaking the symmetry between the points 1 and 2). We shall derive
here two Propositions, based on this idea, whose implementation in
practical computations constitutes a very efficient mean to determine
the partie finie, without any {\it a priori} restriction on the form
of integrand as in the application of the Riesz formula (\ref{33}).

As a matter of fact, in the first proposition, the computation of a
partie-finie integral with two singularities 1 and 2 boils down to the
computation of a partie-finie integral with singularity 1 and a {\it
finite-part} integral (${\rm FP}$) whose singularity is located at
infinity: $r_1\equiv |{\bf x}-{\bf y}_1|\to +\infty$ (so to speak, the
singularity 2 is ``rejected'' to infinity).

\begin{proposition}
For any function $F$ in the class ${\cal F}$ we can write :

\begin{equation}\label{39}
{\rm Pf}_{s_1, s_2} \int d^3{\bf x}~F= {\rm Pf}_{s_1}
\Biggl\{ \sous{\!\!\!\!B\to 0}{\rm FP} \int d^3{\bf r}_1 ~\left( \frac{r_1}{s_2}
\right)^{\!B} \biggl[F -\sum_{b+3 \leq 0} r_2^b  \! \! \! 
\sous{2}{f}_b\biggr] 
\Biggr\}  \; , 
\end{equation}
where the ${}_2f_b$'s denote the coefficients of the expansion of $F$
near $r_2=0$.
\end{proposition}
In words, in order to compute the partie finie one can (i)
``regularize'' $F$ around the point 2 by subtracting out from it the
terms yielding a divergence at 2, i.e.

\begin{equation}\label{40}
{\widetilde F}_2 \equiv F-\sum_{b+3 \leq 0} r_2^b
\! \! \sous{2}{f}_b\;,
\end{equation} 
and (ii) compute the integral of the regularized ${\widetilde F}_2$
using the partie finie around 1 and the finite part when $B\to 0$ to
deal with the divergency at infinity. Notice that the latter
divergency has been introduced simply because of the term
corresponding to $b=-3$ in (\ref{40}) if non-zero.  By finite part
when $B\to 0$ we mean the zeroth-order coefficient in the Laurent
expansion of the analytic continuation with respect to the parameter
$B\in {\mathbb C}$. The analytic continuation is straightforwardly
defined from the domain of the complex plane $\Re (B)>0$ in which the
integral converges at infinity.

\bigskip\noindent
{\it Proof.}  We consider two open domains ${\cal D}_1$ and ${\cal
D}_2$ that are supposed to be disjoined, ${\cal D}_1\bigcap {\cal
D}_2=\emptyset$, complementary in ${\mathbb R}^3$,
i.e. $\overline{{\cal D}_1\bigcup {\cal D}_2}={\mathbb R}^3$, and such
that ${\bf y}_1\in {\cal D}_1$ and ${\bf y}_2\in {\cal D}_2$. From
Definition 4, the partie-finie integral over ${\cal D}_2$ reads as
(for small enough $s$) $$ {\rm Pf} \int_{{\cal D}_2} d^3{\bf
x}~F=\lim_{s\to 0}~\biggl\{\int_{{\cal D}_2\setminus {\cal B}_2(s)}
d^3{\bf x}~F+\sum_{b+3<0}{s^{b+3}\over b+3}\int d\Omega_2
\!\!\!\sous{2}{f}_b +\ln\left({s\over s_2}\right)\int d\Omega_2
\!\!\!\sous{2}{f}_{-3}\biggr\}\;.  $$ Now, two short computations
reveal that

\begin{mathletters}\label{41}\begin{eqnarray}
\sum_{b+3<0} \int_{{\mathbb R}^3\setminus {\cal B}_2(s)} d^3{\bf x}~ r_2^b \! \! \!
\sous{2}{f}_{b} &=& -\sum_{b+3<0} \frac{s^{b+3}}{b+3} \int d\Omega_2  \! \! \!
\sous{2}{f}_{b} \;,\label{41a}\\  
\sous{\!\!\!\!B\to 0}{\rm FP} \int_{{\mathbb R}^3\setminus {\cal B}_2(s)} 
d^3{\bf x} \left(
\frac{|{\bf x}|}{s_2} \right)^B \frac{1}{r_2^3} \! \! \!
\sous{2}{f}_{-3} &=& - \ln \left( \frac{s}{s_2} \right) \int d\Omega_2 \! \! \!
\sous{2}{f}_{-3}\;.\label{41b}
\end{eqnarray}\end{mathletters}$\!\!$
Furthermore, since the integral appearing in (\ref{41a}) is convergent
at infinity, one can add without harm the same finite part operation
when $B\to 0$ as in (\ref{41b}). Thus, the integral over ${\cal D}_2$
may be re-written as

\begin{eqnarray*}
&&\lim_{s\to 0}~\Biggl\{\int_{{\cal D}_2\setminus {\cal B}_2(s)}
d^3{\bf x}~F-\sum_{b+3\leq 0} \sous{\!\!\!\!B\to 0}{\rm FP}
\int_{{\mathbb R}^3\setminus {\cal B}_2(s)} d^3{\bf
x}~\left(\frac{|{\bf x}|}{s_2} \right)^B r_2^b \! \! \!
\sous{2}{f}_{b} 
\Biggr\}\\
&&\qquad\qquad = \lim_{s\to 0}~\Biggl\{\sous{\!\!\!\!B\to 0}{\rm FP}
\int_{{\cal D}_2\setminus {\cal B}_2(s)} d^3{\bf x}~\left(\frac{|{\bf
x}|}{s_2} \right)^B {\widetilde F}_2 - \sum_{b+3\leq 0}
\sous{\!\!\!\!B\to 0}{\rm FP} \int_{{\cal D}_1} d^3{\bf
x}~\left(\frac{|{\bf x}|}{s_2} \right)^B r_2^b \! \! \!
\sous{2}{f}_{b}\Biggl\}\\
&&\qquad\qquad = \lim_{s\to 0}~\Biggl\{\sous{\!\!\!\!B\to 0}{\rm FP} 
\int_{{\cal D}_2} d^3{\bf x}~\left(\frac{|{\bf x}|}{s_2} \right)^B {\widetilde F}_2 
- \sum_{b+3\leq 0} \sous{\!\!\!\!B\to 0}{\rm FP} \int_{{\cal D}_1\setminus 
{\cal B}_1(s)} d^3{\bf x}~\left(\frac{|{\bf x}|}{s_2} \right)^B r_2^b \! \! \!
\sous{2}{f}_{b}\Biggl\}\;.
\end{eqnarray*}
We have used the facts that the integral of $F$ converges at infinity
(first equality) and the integral of ${\widetilde F}_2$ converges at
the singularity 2 (second equality).  Adding up the other contribution
extending over ${\cal D}_1$, we readily obtain the complete partie
finie as $$
\lim_{s \to 0} \left\{\sous{\!\!\!\!B\to 0}{\rm FP} 
\int_{{\mathbb R}^3\setminus {\cal B}_1(s)} d^3{\bf x} \left(
\frac{|{\bf x}|}{s_2} \right)^B {\widetilde F}_2 +
\sum_{a+3<0} \frac{s^{a+3}}{a+3} \int d\Omega_1 \! \! \!
\sous{1}{f}_{a} + \ln \left(
\frac{s}{s_1} 
\right) \int d\Omega_1 
\! \! \! \sous{1}{f}_{-3} \right\}\;.
$$ Since the coefficients ${}_1f_a$, for $a\leq -3$, are those of the
expansion when $r_1\to 0$ of $F$ as well as of ${\widetilde F}_2$, we
recognize in the expression above the partie finie (with respect to 1
only) of the integral of the regularized function ${\widetilde
F}_2$. Hence the intermediate expression

\begin{equation}\label{42}
{\rm Pf}_{s_1, s_2} \int d^3{\bf x}~F= {\rm Pf}_{s_1}
\Biggl\{ \sous{\!\!\!\!B\to 0}{\rm FP} \int d^3{\bf x} ~\left( 
\frac{|{\bf x}|}{s_2}
\right)^{\!B} {\widetilde F}_2\Biggr\} \;.
\end{equation}
To establish the proposition it remains to change of variable ${\bf
x}$ into ${\bf r}_1$. At that point, we must be careful, because under
this change of variable the regularization factor $|{\bf x}|^B$
changes itself in a complicated way. Fortunately, we can limit
ourselves to the case where $B$ is infinitesimal, since we shall take
the finite part afterwards, making $B\to 0$. We substitute to $|{\bf
x}|^B$ in the right side of (\ref{42}) its equivalent expression in
terms of ${\bf r}_1$ and where we expand when $B\to 0$, i.e.

\begin{equation}\label{43}
|{\bf x}|^B=r_1^B e^{B\ln\Bigl(\case{|{\bf
x}|}{r_1}\Bigr)}=r_1^B\Biggl\{1+{B\over 2}\ln \left[1+2\case{{\bf
n}_1.{\bf y}_1}{r_1}+\case{{\bf
y}_1^2}{r_1^2}\right]+O(B^2)\Biggr\}\;,
\end{equation}
where ${\bf n}_1.{\bf y}_1$ denotes the usual scalar product on
${\mathbb R}^3$ (and ${\bf y}_1^2={\bf y}_1.{\bf y}_1$). Now, the
dominant term in the latter expansion amounts simply to replacing
$|{\bf x}|^B$ by $r_1^B$, which would yield precisely the result
(\ref{39}) we want to prove; but we have still to show that all the
extra terms in the expansion (\ref{43}), which carry at least a factor
$B$ in front, do not contribute to the final result, i.e. that

\begin{equation}\label{44} 
\sous{\!\!\!\!B\to 0}{\rm FP} \Biggl[{B\over 2}\int^{+\infty} 
d^3{\bf r}_1 ~\left( \frac{r_1}{s_2}
\right)^{\!B} \ln \left[1+2\case{{\bf n}_1.{\bf y}_1}{r_1}
+\case{{\bf y}_1^2}{r_1^2}\right] ~{\widetilde F}_2 + O(B^2) \Biggr] = 0\;.
\end{equation}
Because of the factor $B$ in front, the only possible contribution to
the finite part for $B\to 0$ occurs when the integral develops a pole
at $B=0$ due to the behaviour of the integrand at infinity ($r_1\to
+\infty$). Hence, as indicated in (\ref{44}), the value of the
integral depends only on the bound at infinity [this is also why we
did not write a ${\rm Pf}_{s_1}$ symbol in (\ref{44}): the partie
finie deals with the bound $r_1=0$, which is irrelevant to this case].
In order to evaluate the pole, we replace the integrand by its
expansion when $r_1\to +\infty$. We know that $F$ behaves as
$o\Bigl(\case{1}{|{\bf x}|^3}\Bigr)$ at a maximum $|{\bf
x}|\to+\infty$ to ensure the convergence of the integral of $F$ at
infinity, swhen o we have $F=o\Bigl(\case{1}{r_1^3}\Bigr)$ when
$r_1\to+\infty$. Now, from the defining expression (\ref{40}) of
${\widetilde F}_2$, we obtain

\begin{equation}\label{45}
{\widetilde F}_2=-{1\over r_1^3}\!\sous{2}{f}_{-3}({\bf
n}_1)+o\left({1\over r_1^3}\right)\qquad\hbox{when $~r_1\to
+\infty$}\;,
\end{equation}
after making the replacements of $r_2$ and ${\bf n}_2$ by $r_1$ and
${\bf n}_1$ which are permitted because we are working at the dominant
order when $r_1\to +\infty$.  On the other hand, we have $\ln
\left[1+2\case{{\bf n}_1.{\bf y}_1}{r_1}+\case{{\bf
y}_1^2}{r_1^2}\right] =2\case{{\bf n}_1.{\bf
y}_1}{r_1}+O\Bigl(\case{1}{r_1^2}\Bigr)$. So that the integral to be
computed (as concerns the only relevant bound at infinity) reads as $$
\int^{+\infty} d^3{\bf r}_1 ~r_1^B \ln \left[1+2\case{{\bf n}_1.{\bf y}_1}{r_1}
+\case{{\bf y}_1^2}{r_1^2}\right] ~{\widetilde F}_2 = -2 \int^{+\infty} d r_1 
~r_1^{B-2} \biggl\{ \int d\Omega_1 
~{\bf n}_1.{\bf y}_1 \!\!\sous{2}{f}_{-3}({\bf n}_1) +
o\left(r_1^0\right) \biggr\}\;.  $$ This integral cannot generate a
pole at $B=0$ since such a pole could come only from a radial integral
of the type $\int^{+\infty} dr_1 r_1^{B-1}$ (after the angular
integration has been performed). Repeating the same reasoning to any
higher orders in $B$, we prove the equation (\ref{44}) as well as
Proposition 1.

In practice, Proposition 1 is used with the integration with respect
to ${\bf n}_1$, followed by the integration over $r_1$ varying from 0
(${\rm Pf}_{s_1}$ takes care of this bound) to infinity (where ${\rm
FP}_{B\to 0}$ does the work); Proposition 1 justifies this process
even when the original integral is divergent at {\it both}
singularities.  The result of the angular integration depends on where
the field point is located, either inside the ball ${\cal
B}_1(r_{12})$ centered on ${\bf y}_1$ and of radius $r_{12}$ (the
point 2 lies on the surface of this ball) or in the complementary
domain ${\mathbb R}^3\setminus {\cal B}_1(r_{12})$. Therefore, a
natural splitting of the integral (\ref{39}) is

\begin{equation}\label{46}
{\rm Pf}_{s_1, s_2} \int d^3{\bf x}~F=~{\rm Pf}_{s_1} \int_{{\cal
B}_1(r_{12})} d^3{\bf r}_1 ~{\widetilde F}_2 ~+~
\sous{\!\!\!\!B\to 0}{\rm FP} \int_{{\mathbb R}^3\setminus 
{\cal B}_1(r_{12})} d^3{\bf r}_1 ~\left( \frac{r_1}{s_2}
\right)^{\!B} {\widetilde F}_2 \; ,
\end{equation}
taking into account the fact that the partie finie ${\rm Pf}_{s_1}$
applies only to the inner integral, over ${\cal B}_1(r_{12})$, and the
finite part ${\rm FP}_{B\to 0}$ only to the outer one, over ${\mathbb
R}^3\setminus {\cal B}_1(r_{12})$. To be more specific, the angular
integral of ${\widetilde F}_2$ defines two angular-average functions
${\widetilde I}_2(r_1)$ and ${\widetilde J}_2(r_1)$ depending on
whether ${\bf x}$ is in ${\cal B}_1(r_{12})$ or its complementary:

\begin{equation}\label{46'}
\int {d\Omega_1\over 4\pi}~{\widetilde F}_2=\left\{
\vcenter{\vskip 0.1pc\hbox{$~{\widetilde I}_2(r_1)\qquad$ 
when $r_1\leq r_{12}~,$}\vskip 0.5pc\hbox{
${\widetilde J}_2(r_1)\qquad$ when $r_1>r_{12}~.$}\vskip 0.1pc}
\right.
\end{equation}
The functions ${\widetilde I}_2$ and ${\widetilde J}_2$ depend also
explicitely on the source points ${\bf y}_1$ and ${\bf y}_2$. [As an
example, in the case ${\widetilde F}_2=r_2$, we find ${\widetilde
I}_2=r_{12}+{r_1^2\over 3r_{12}}$ and ${\widetilde
J}_2=r_1+{r_{12}^2\over 3r_1}$.]  Now, knowing ${\widetilde I}_2$ and
${\widetilde J}_2$, we can achieve the radial integration according to
the formula

\begin{equation}\label{46''}
{\rm Pf}_{s_1, s_2} \int d^3{\bf x}~F=4\pi~\! {\rm Pf}_{s_1} 
\int_0^{r_{12}} d r_1 ~r_1^2~\!{\widetilde I}_2 ~+~4\pi
\sous{\!\!\!\!B\to 0}{\rm FP} \int_{r_{12}}^{+\infty} d r_1 \left( 
\frac{r_1}{s_2}
\right)^{\!B} r_1^2~\!{\widetilde J}_2 \;. 
\end{equation}
The first term in (\ref{46''}) is quite simple to handle in practice,
whereas the second one is more difficult because it requires {\it a
priori} the knowledge of a closed-form expression for the integral of
$r_1^{B+2}~\!{\widetilde J}_2$, valid for any $B$ such that $\Re
(B)>0$. Obtaining this may not be feasible if $F$ is too complicated;
in this event, we should use a different form of the integral at
infinity. The second proposition, which provides the appropriate form,
constitutes, perhaps, the most powerful way to compute the partie
finie in rather complicated applications.

\begin{proposition}
The partie finie of the integral of $F\in {\cal F}$ (if convergent at
infinity) reads as:

\begin{eqnarray}\label{47}
{\rm Pf}_{s_1, s_2} \int d^3{\bf x}~F&=&4\pi~\! {\rm Pf}_{s_1}
\int_0^{r_{12}} d r_1 ~r_1^2~\!{\widetilde I}_2(r_1) ~+~4\pi
\int_{r_{12}}^{+\infty} d r_1~\biggl[r_1^2~\!{\widetilde J}_2(r_1)+{1\over r_1}
\left(r_2^3 F\right)_2\biggr]\nonumber\\
~~&+&4\pi \left(r_2^3 F\right)_2 \ln\left({r_{12}\over s_2}\right)\;,
\end{eqnarray}
(and similarly by interchange of $1$ and $2$).
\end{proposition}
{\it Proof.} Consider the angular average of the expansion of
${\widetilde F}_2$ when $r_1\to +\infty$ which has been determined in
(\ref{45}). We get

\begin{equation}\label{47'}
{\widetilde J}_2 \equiv \int {d\Omega_1\over 4\pi}~{\widetilde F}_2 =
- {1\over r_1^3} ~\bigl(r_2^3 F\bigr)_2 + o\left({1\over r_1^3}\right)
\;,
\end{equation}
where the coefficient of the dominant term is made of a Hadamard
partie finie at point 2.  Let us subtract and add to ${\widetilde
J}_2$ inside the second integral in (\ref{46''}) the previous dominant
term at infinity. In this way, we may re-write it as the sum of a
convergent integral at infinity on one hand, to which we can then
remove the finite part prescription, and a simple extra integral on
the other hand. Namely,

\begin{eqnarray*}
\int_{r_{12}}^{+\infty} d r_1 ~\biggl[r_1^2~\!{\widetilde J}_2
+{1\over r_1}\left(r_2^3 F\right)_2\biggr]~&-&~\bigl(r_2^3 F\bigr)_2 
~\sous{\!\!\!\!B\to 0}{\rm FP} \int_{r_{12}}^{+\infty} {d r_1\over r_1}
~\biggl( \frac{r_1}{s_2}
\biggr)^{\!B}\;.
\end{eqnarray*}
The extra integral is finally computed in a simple way as

\begin{eqnarray*}
\sous{\!\!\!\!B\to 0}{\rm FP} \int_{r_{12}}^{+\infty} {d r_1\over r_1}
~\biggl( \frac{r_1}{s_2}
\biggr)^{\!B}=\sous{\!\!\!\!B\to 0}{\rm FP} \Biggl[-{1\over B}
\biggl(\frac{r_{12}}{s_2}
\biggr)^{\!B}\Biggr]=-\ln\left({r_{12}\over s_2}\right)\;,
\end{eqnarray*}
where we used the properties of the analytic continuation. {\it QED}.

Thanks to Proposition 2 we are now able to compute many integrals
which could not be deduced from the Riesz formula (\ref{33}), unlike
for (\ref{37}). For instance,

\begin{mathletters}\label{48}\begin{eqnarray}
&&{\rm Pf}_{s_1,s_2}\int {d^3{\bf x}\over
r_1^3~\!r_2^3~\!(r_1+r_2)}={4\pi\over
r_{12}^4}\Biggl[\ln\left({r_{12}\over
s_1}\right)+\ln\left({r_{12}\over s_2}\right)-{8\over 3}\ln 2+{2\over
3}\Biggr]\;,\label{48a}\\ &&{\rm Pf}_{s_1,s_2}\int {d^3{\bf x}\over
r_1^3~\!r_2^3~\!(r_1+r_2+r_{12})}={2\pi\over
r_{12}^4}\Biggl[\ln\left({r_{12}\over
s_1}\right)+\ln\left({r_{12}\over s_2}\right)+{\pi^2\over
\!3}-4\Biggr]\;.\label{48b}
\end{eqnarray}\end{mathletters}$\!\!$ 
The result for the integral (\ref{48b}) is in agreement with the one
that follows from a recent generalization of the Riesz formula to
include arbitrary powers of $r_1+r_2+r_{12}$, which has been obtained
by Jaranowski and Sch\"afer (see the appendix B.2 in
\cite{JaraS98}). In any case, the dependence of the partie-finie
integral on the two constants $s_1$ and $s_2$ is given by

\begin{eqnarray}\label{48'}
{\rm Pf}_{s_1,s_2} \int d^3{\bf x}~F &=& 4\pi \left(r_1^3 F\right)_1
\ln\left({r_{12}\over s_1}\right) +4\pi \left(r_2^3 F\right)_2
\ln\left({r_{12}\over s_2}\right)\nonumber\\ &+&\hbox{terms
independent of $s_1$, $s_2$}\;.
\end{eqnarray}

\section{Partie finie of Poisson integrals}

In this section we investigate the main properties of the partie finie
of Poisson integrals of singular functions in the class ${\cal F}$. We
have in view the application to the post-Newtonian motion of particles
in general relativity, since the post-Newtonian iteration proceeds
typically through Poisson (or Poisson-type) integrals. Consider a
fixed (``spectator'') point ${\bf x}'\in {\mathbb R}^3$ and, for each
value of ${\bf x}'$, define the function $S_{{\bf x}'}({\bf x})=F({\bf
x})/|{\bf x}-{\bf x}'|$ where $F\in {\cal F}$. Clearly, for any given
${\bf x}'$, the function $S_{{\bf x}'}$ belongs to the class ${\cal
F}_{\rm loc}$, introduced in Section II, Definition 2. In addition,
when the spectator point ${\bf x}'$ coincides with the singular point
${\bf y}_1$ (and similarly for ${\bf y}_2$), we have $S_{{\bf y}_1}\in
{\cal F}$. Since (as already mentionned) Definition 4 can be extended
to functions in the class ${\cal F}_{\rm loc}$, we can consider the
partie-finie integral

\begin{equation}\label{49}
P({\bf x}')=-{1\over 4\pi}~{\rm Pf}\int d^3{\bf x}~\!S_{{\bf x}'}({\bf
x})=-{1\over 4\pi} ~{\rm Pf} \int \frac{d^3{\bf x}}{|{\bf x}-{\bf
x}'|} F({\bf x}) \; .
\end{equation} 
This is, indeed, what we shall call the ``Poisson'' integral of
$F$. In particular, when the spectator point ${\bf x}'$ is equal to
${\bf y}_1$, we shall write

\begin{equation}\label{49'}
P({\bf y}_1)=-{1\over 4\pi}~{\rm Pf}\int d^3{\bf x}~\!S_{{\bf
y}_1}({\bf x})=-{1\over 4\pi} ~{\rm Pf} \int \frac{d^3{\bf x}}{r_1}
F({\bf x}) \; .
\end{equation} 
The Poisson integral is not continuous at the singular point ${\bf
y}_1$ because $P({\bf x}')$, when initially defined for ${\bf x}'\not=
{\bf y}_1$, admits an expansion that is singular when ${\bf x}'$ tends
to ${\bf y}_1$. In the present Section, our aim is to understand the
limit relation of the integral $P({\bf x}')$ when $r'_1\equiv |{\bf
x}'-{\bf y}_1|\to 0$, and to connect it with the ``regularized''
integral $P({\bf y}_1)$ given by (\ref{49'}). In particular, we shall
show that the ``partie finie'' (in an extended Hadamard's sense) of
$P({\bf x}')$ at ${\bf x}'={\bf y}_1$ is related in a precise way to
$P({\bf y}_1)$. Let us make clear straight away that $P({\bf x}')$, as
a function of ${\bf x}'$ different from ${\bf y}_1$ (and ${\bf y}_2$),
does not belong to the class ${\cal F}$ as the Poisson integral
typically generates logarithms in the expansion when $r'_1\to 0$.  In
particular, the coefficient of zeroth power of $r'_1$ in the latter
expansion contains {\it a priori} a $\ln r'_1$ term, and its partie
finie in the sense of Definition 3 is in fact not finite at all,
because of the presence of this formally infinite constant $\ln
r'_1=-\infty$. A possible way to deal with this problem, followed by
Sellier in \cite{Sellier}, is to {\it exclude} the $\ln r'_1$ (and any
higher power of $\ln r'_1$) from the definition of the partie
finie. On the other hand, in applications to the physical problem, the
constant $\ln r'_1$ can be viewed as a ``renormalization'' constant,
which is better to keep as it appears all the way through the
calculation. Therefore, we simply include here the renormalization
constant $\ln r'_1$ into the definition; but, for simplicity's sake,
we stick to the name of ``partie finie'' in this case (although the
$\ln r'_1$ makes it formally infinite). Thus, for a function like $P$
admitting a logarithmic expansion:

\begin{equation}\label{50}
\forall N\in{\mathbb N}\;,\quad P({\bf x}')=\sum_{a\leq N\atop p=0,1} 
{r'_1}^a (\ln r'_1)^p \!\!\sous{1}{f}_{a,p}({\bf n}'_1) + o\Bigl({r'_1}^N\Bigr)
\quad\mbox{when $~r'_1\to 0$} \;,
\end{equation}
we define the Hadamard partie finie of $P$ at 1 by
 
\begin{equation}\label{51}
(P)_1 = \int {d\Omega'_1\over 4\pi} ~\Bigl[\!\!\sous{1}{f}_{0,0}({\bf
n}'_1)+\!\!\sous{1}{f}_{0,1}({\bf n}'_1)\ln r'_1\Bigr]\;.
\end{equation}

\begin{theorem}
The Hadamard partie finie at 1 (in the previous sense) of the Poisson
integral of any $F\in {\cal F}$ reads as

\begin{equation}\label{52}
(P)_1=-{1\over 4\pi} ~{\rm Pf}_{s_1, s_2} \int \frac{d^3{\bf x}}{r_1}
~F({\bf x})+\Biggl[\ln\left({r_1'\over s_1}\right)-1\Biggr]
\bigl(r_1^2 F\bigr)_1 \;,
\end{equation}
with $r_1'=|{\bf x}'-{\bf y}_1|$. Furthermore the constants $s_1$
cancel each other from the two terms in the right side of (\ref{52})
(so the partie finie depends on the two constants $\ln r'_1$ and $\ln
s_2$).
\end{theorem}
In words, the partie finie of the Poisson integral at 1 is equal to
the regularized integral $P({\bf y}_1)$, obtained from the replacement
${\bf x}'\rightarrow {\bf y}_1$ inside the integrand of $P({\bf x}')$,
augmented by a term associated with the presence of the (infinite)
constant $\ln r'_1$.

\bigskip\noindent
{\it Proof.} The fact that the constants $s_1$ cancel out (so $s_1$ is
``replaced'' by $r'_1$) is a trivial consequence of the dependence of
the partie finie on $s_1$ and $s_2$ determined in (\ref{48'}). For our
proof, we need the explicit expressions of the objects $P({\bf x}')$,
when ${\bf x}'$ is different from ${\bf y}_1$ and ${\bf y}_2$, and
$P({\bf y}_1)$, following from Definition 4. For ${\bf x}'\not= {\bf
y}_1$ and $r_1=|{\bf x}-{\bf y}_1|\to 0$, we have the expansion

\begin{equation}\label{53}
S_{{\bf x}'}({\bf x})= \sum_{l \ge 0} \frac{(-)^l}{l!}
\partial_L' \!\!\left({1\over r_1'}\right) \sum_a
r_1^{a+l} n_1^L \! \! \! \! \sous{1}{f}_a({\bf n}_1)
\end{equation}
(and {\it idem} $1\leftrightarrow 2$), where $r_1'=|{\bf x}'-{\bf
y}_1|$, $\partial_L'$ being the multi-spatial derivative acting on
${\bf x}'$. From (\ref{21''}), we get the expression (for ${\bf
x}'\not= {\bf y}_1$ and ${\bf y}_2$)

\begin{eqnarray}\label{53'}
P({\bf x}')=&-&{1\over 4\pi} \lim_{s\to 0} \Biggl\{ \int_{{\mathbb
R}^3\setminus {\cal B}_1(s)\cup {\cal B}_2(s)}
\frac{d^3{\bf x}}{|{\bf x}-{\bf x}'|}~F\nonumber\\
&+&\sum_{l \ge 0} \frac{(-)^l}{l!} \partial'_L \!\!\left({1\over
r_1'}\right)
\biggl[ \sum_{a+l+3<0} 
 \frac{s^{a+l+3}}{a+l+3}
\int d\Omega_1 ~n_1^L \! \! \! \!
\sous{1}{f}_a +\ln\left({s\over s_1}\right)\int d\Omega_1 ~n_1^L \! \! \! \!
\sous{1}{f}_{-3-l} \biggr]\nonumber\\
&+&1\leftrightarrow 2\}\;.
\end{eqnarray}
Applying the recipe (\ref{51}), we start by computing the angular
integral over ${\bf n}'_1=({\bf x}'-{\bf y}_1)/r'_1$ (for a fixed
$r'_1$) of $P({\bf x}')$ in the form given by (\ref{53'}), and
consider the limit $r'_1\to 0$ afterwards. Since $s$ is fated to tend
to zero first, one can choose $s<r'_1$, and as we are ultimately
interested in the limit $r'_1\to 0$, we also assume $r'_1<r_{12}$.  To
compute the angular average of the divergent terms in (\ref{53'}), we
make use of the identities

\begin{mathletters}\label{54}\begin{eqnarray}
\int {d\Omega'_1\over 4\pi}~\partial'_L \!\!\left({1\over r_1'}\right) 
&=& {\delta_{0l}\over r'_1}\;,\\
\int {d\Omega'_1\over 4\pi}~\partial'_L \!\!\left({1\over r_2'}\right) 
&=& \partial_L \!\!\left({1\over r_{12}}\right) 
\end{eqnarray}\end{mathletters}$\!\!$
(where $\delta_{0l}$ denotes the Kronecker symbol). On the other
hand, the relevant formula to treat the integral in the right side of
(\ref{53'}) is

\begin{equation}\label{55}
\int {d\Omega'_1\over 4\pi}~{1\over |{\bf x}-{\bf x}'|}=\left\{
\vcenter{\vskip 0.1pc\hbox{$~{1\over r'_1}\qquad$ (if $r_1<r'_1$)~,}
\vskip 0.5pc\hbox{
${1\over r_1}\qquad$ (if $r'_1<r_1$)~.}\vskip 0.1pc}
\right.
\end{equation}
We split this integral into three other ones, the first of them
extending over the ``exterior'' domain ${\mathbb R}^3\setminus {\cal
B}_1(r'_1)\cup {\cal B}_2(r'_1)$, and the two remaining ones over the
ring-shaped regions ${\cal B}_1(r'_1)\setminus {\cal B}_1(s)$ and
${\cal B}_2(r'_1)\setminus {\cal B}_2(s)$. Hence:

\begin{eqnarray}\label{56}
\int {d\Omega'_1\over 4\pi}P({\bf x}')&=&-{1\over 4\pi} \lim_{s\to 0} 
\Biggl\{\int_{{\mathbb R}^3\setminus {\cal B}_1(r'_1)\cup {\cal B}_2
(r'_1)}\frac{d^3{\bf x}}{r_1}~F
\nonumber\\
&+&{1\over r'_1} \int_{{\cal B}_1(r'_1)\setminus {\cal B}_1(s)}
d^3{\bf x}~F +\int_{{\cal B}_2(r'_1)\setminus {\cal
B}_2(s)}\frac{d^3{\bf x}}{r_1}~F\nonumber\\ &+&{1\over
r'_1}\biggl[\sum_{a+3<0} \frac{s^{a+3}}{a+3}
\int d\Omega_1 \! \! \! \!
\sous{1}{f}_a +\ln\left({s\over s_1}\right)\int d\Omega_1 ~\! \! \! \!
\sous{1}{f}_{-3}\biggr]\nonumber\\ 
&+&\sum_{l \ge 0} \frac{(-)^l}{l!} \partial_L \!\!\left({1\over r_{12}}\right) 
\biggl[ ~\sum_{b+l+3<0} 
 \frac{s^{b+l+3}}{b+l+3}
\int d\Omega_2 ~n_2^L \! \! \! \!
\sous{2}{f}_b 
+\ln\left({s\over s_2}\right)\int d\Omega_2 ~n_2^L \! \! \! \!
\sous{2}{f}_{-3-l} \biggr]\Biggr\}\;.\nonumber\\
\end{eqnarray}
Next, supposing that $r'_1$ is small enough, we may replace $F$ in the
second and third terms by its own expansions around 1 and 2
respectively. We find that the divergent terms in $s$ cancel out, so
we are allowed to apply the limit $s\to 0$. This yields

\begin{eqnarray}\label{57}
\int {d\Omega'_1\over 4\pi}P({\bf x}')&=&-{1\over 4\pi} \Biggl\{\int_
{{\mathbb R}^3\setminus {\cal B}_1(r'_1)\cup {\cal B}_2(r'_1)}\frac{d^3
{\bf x}}{r_1}~F
\nonumber\\
&+&{1\over r'_1}\biggl[\sum_{a+3<0} \frac{{r'_1}^{a+3}}{a+3}
\int d\Omega_1 \! \! \! \!
\sous{1}{f}_a +\ln\left({r'_1\over s_1}\right)\int d\Omega_1 ~\! \! \! \!
\sous{1}{f}_{-3}+r'_1\int d\Omega_1 \! \! \! \!
\sous{1}{f}_{-2} \biggr]\nonumber\\ 
&+&\sum_{l \ge 0} \frac{(-)^l}{l!} \partial_L \!\!\left({1\over r_{12}}\right) 
\biggl[ ~\sum_{b+l+3<0} 
 \frac{{r'_1}^{b+l+3}}{b+l+3}
\int d\Omega_2 ~n_2^L \! \! \! \!
\sous{2}{f}_b 
+\ln\left({r'_1\over s_2}\right)\int d\Omega_2 ~n_2^L \! \! \! \!
\sous{2}{f}_{-3-l} \biggr]\nonumber\\
&+&o({r'_1}^0)\}
\end{eqnarray}
(the remainder dies out when $r'_1\to 0$). Under the latter form we
recognize most of the terms composing the integral $P({\bf
y}_1)$. Indeed, we have, respectively when $r_1\to 0$ and $r_2\to 0$,

\begin{mathletters}\label{57'}\begin{eqnarray}
S_{{\bf y}_1}({\bf x})&=& \sum_a r_1^{a-1} \! \! \! \!
\sous{1}{f}_a({\bf n}_1)\label{57'a}\;,\\ S_{{\bf y}_1}({\bf x})&=&
\sum_{l \ge 0} \frac{(-)^l}{l!}
\partial_L\!\!\left({1\over r_{12}}\right) \sum_b
r_2^{b+l} n_2^L \! \! \! \! \sous{2}{f}_b({\bf n}_2)\;.\label{57'b}
\end{eqnarray}\end{mathletters}$\!\!$
Now, using the form (\ref{22'}) of the partie finie with the change of
notation $s'=r'_1$, we find

\begin{eqnarray}\label{58}
P({\bf y}_1)=&-&{1\over 4\pi}\Biggl\{ \int_{{\mathbb R}^3\setminus
{\cal B}_1(r'_1)\cup {\cal B}_2(r'_1)}
\frac{d^3{\bf x}}{r_1}~F+\sum_{a+2<0} 
 \frac{{r'_1}^{a+2}}{a+2}
\int d\Omega_1 \! \! \! \!
\sous{1}{f}_a +\ln\left({r'_1\over s_1}\right)\int d\Omega_1 ~\! \! \! \!
\sous{1}{f}_{-2}\nonumber\\ 
&+&\sum_{l \ge 0} \frac{(-)^l}{l!} \partial_L \!\!\left({1\over r_{12}}\right) 
\biggl[ ~\sum_{b+l+3<0} 
 \frac{{r'_1}^{b+l+3}}{b+l+3}
\int d\Omega_2 ~n_2^L \! \! \! \!
\sous{2}{f}_b 
+\ln\left({r'_1\over s_2}\right)\int d\Omega_2 ~n_2^L \! \! \! \!
\sous{2}{f}_{-3-l} \biggr]\nonumber\\
&+&o({r'_1}^0)\}\;.
\end{eqnarray}
We finally evaluate the difference between (\ref{57}) and (\ref{58})
and look for the partie finie in the sense of (\ref{51}) (i.e. keeping
the $\ln r'_1$ term). We obtain

\begin{equation}\label{58'}
(P)_1-P({\bf y}_1)=\Biggl[\ln\left({r_1'\over s_1}\right)-1\Biggr]\int
{d\Omega_1\over 4\pi}
\! \! \!\sous{1}{f}_{-2}\qquad\hbox{({\it QED}).}
\end{equation}

The same type of result can be proved for the partie finie of the
``twice-iterated'' Poisson integral defined by

\begin{equation}\label{59}
Q({\bf x}')=-{1\over 4\pi} ~{\rm Pf} \int d^3{\bf x} ~|{\bf x}-{\bf x}'|
F({\bf x}) \; .
\end{equation}
We find, analogously to (\ref{52}), that

\begin{equation}\label{60}
(Q)_1=-{1\over 4\pi} ~{\rm Pf} \int d^3{\bf x}~r_1
F({\bf x})+\Biggl[\ln\left({r_1'\over s_1}\right)+{1\over 2}\Biggr] 
\bigl(r_1^4 F\bigr)_1\;.
\end{equation}
For the parties finies of the {\it gradients} of the Poisson and
twice-iterated Poisson integrals, we get

\begin{mathletters}\label{61}\begin{eqnarray}
(\partial_iP)_1&=&-{1\over 4\pi} ~{\rm Pf} \int d^3{\bf x}~{n_1^i\over
r_1^2} ~F({\bf x})+\ln\left({r_1'\over s_1}\right) \bigl(n_1^i r_1
F\bigr)_1 \;,\\ (\partial_iQ)_1&=&{1\over 4\pi} ~{\rm Pf} \int d^3{\bf
x}~n_1^i F({\bf x})-\Biggl[\ln\left({r_1'\over s_1}\right)-{1\over
2}\Biggr] \bigl(n_1^i r_1^3 F\bigr)_1 \;.
\end{eqnarray}\end{mathletters}$\!\!$
Those results are proved in the same way as in Theorem 3 (with similar
cancellations of the constants $s_1$).

\section{Partie finie pseudo-functions}

\subsection{A class of pseudo-functions}

The concept of Hadamard partie finie of the divergent integral of
functions $F\in {\cal F}$ yields a natural definition of a class of
pseudo-functions ${\rm Pf}F$ (``partie finie'' of $F$), namely linear
forms on a subset of ${\cal F}$, of the type $G\in {\cal F}\rightarrow
<{\rm Pf}F,G>\in {\mathbb R}$, where the result of the action of ${\rm
Pf}F$ on $G$ is denoted using a duality bracket $<,\!>$.

\begin{definition}
For any function $F\in {\cal F}$ we define the pseudo-function ${\rm
Pf}F$ as the linear functional which associates to any $G\in {\cal
F}$, such that $FG=o(|{\bf x}|^{-3})$ when $|{\bf x}|\to +\infty$, the
partie-finie integral of the product $FG$, i.e.

\begin{equation}\label{62}
<{\rm Pf}F,G>={\rm Pf}\int d^3{\bf x}~FG\;,
\end{equation}
where the partie-finie integral is defined by ${\rm (\ref{21''})}$. 
\end{definition}
As we can see, the pseudo-function ${\rm Pf}F$ is not a linear form on
${\cal F}$ itself but on the subset of ${\cal F}$ such that the
integral converges at infinity. For simplicity's sake we will always
say that statements like (\ref{62}) are valid $\forall G\in {\cal F}$,
without mentioning this restriction. Note also that the partie-finie
integral depends on the two constants $s_1$, $s_2\in {\mathbb
R}^{+*}$, and so is the pseudo-function which should indeed be denoted
${\rm Pf}_{s_1, s_2}F$. In our simplified notation we omit indicating
$s_1$ and $s_2$.

An evident property of the duality bracket is its ``symmetry'' by
exchanging the roles of the two slots of the bracket, namely:

\begin{equation}\label{63}
\forall (F, G)\in {\cal F}^2\;,\quad <{\rm Pf}F,G>=<{\rm Pf}G,F>\;.
\end{equation}
Also evident are the properties: $$ <{\rm Pf}F,GH>=<{\rm
Pf}G,FH>=<{\rm Pf}(FG),H>=<{\rm Pf}(FGH)~,1>\;.  $$ In the following
we generally do not distinguish between the two slots in
$<,\!>$. Accordingly we {\it define} the object $$ <F,{\rm Pf}G>\equiv
<{\rm Pf}G,F>\;.  $$ Even more, we allow for a bracket in which the
two slots are filled with pseudo-functions. Thus, we write $$ <{\rm
Pf}F,{\rm Pf}G>\equiv <{\rm Pf}F,G>=<{\rm Pf}G,F>\;, $$ which
constitutes merely the {\it definition} of the new object $<{\rm
Pf}F,{\rm Pf}G>$.

We denote by ${\cal F}'$ the set of pseudo-functions ${\rm Pf}F$, when
$F$ describes the class ${\cal F}$, introduced by Definition 5~:
${\cal F}'=\Bigl\{ {\rm Pf}F;~F\in {\cal F}\Bigr\}$. Later we shall
extend the definition of ${\cal F}'$ to include the ``limits'' of some
pseudo-functions. Roughly, the set ${\cal F}'$ plays a role analogous
to the set ${\cal D}'$ in distribution theory \cite{Schwartz}, which
is dual to the class ${\cal D}$ of functions which are both
$C^\infty({\mathbb R}^3)$ (for what we are concerned about here) and
zero outside a compact subset of ${\mathbb R}^3$. In distribution
theory the set ${\cal D}$ is endowed with the Schwartz topology~: a
sequence $(\varphi_n)_{n\in {\mathbb N}}$ of elements of ${\cal D}$
converges to zero if and only if (i) $\exists n_0\in {\mathbb N}$ and
a compact $K$ of ${\mathbb R}^3$ such that $\forall n\geq n_0$, ${\rm
supp}(\varphi_n) \subset K$, and (ii) for any multi-index
$L=i_1i_2\cdots i_l$, $\partial_L\varphi_n$ converges uniformly to
zero. ${\cal D}'$ is the set of linear forms on ${\cal D}$ that are
continuous with respect to that topology. In this paper we shall not
attempt to define a topology on the class ${\cal F}$, and shall limit
ourselves (having in view the physical application) to the definition
of the algebraic and differential rules obeyed by the pseudo-functions
of ${\cal F}'$. However we can state~:

\begin{lemma} 
The pseudo-functions of ${\cal F}'$, when restricted to the set ${\cal
D}$ of $C^\infty({\mathbb R}^3)$ functions with compact support, are
distributions in the sense of Schwartz~:

\begin{equation}\label{64}
{\rm Pf}F_{~\bigl|_{\cal D}}\in ~{\cal D}'\;.
\end{equation}
\end{lemma}
{\it Proof}. All we need to check is that the pseudo-function ${\rm
Pf}F_{~\bigl|_{\cal D}}$ is continuous with respect to the Schwartz
topology \cite{Schwartz}. Consider a sequence $\varphi_n\in {\cal D}$
tending to zero in the sense recalled above. Applying the partie-finie
integral in the form (\ref{22'}), we get ($\forall s'\ll 1$ and
$\forall N\in {\mathbb N}$)

\begin{eqnarray*}
<{\rm Pf}F_{~\bigl|_{\cal D}},\varphi_n> &=&\int_{{\mathbb
R}^3\setminus {\cal B}_1(s')\cup {\cal B}_2(s')} d^3{\bf x}~F
\varphi_n\nonumber\\ &+& \sum_{l \ge 0} \frac{1}{l!}
\partial_L\varphi_n({\bf y}_1) \biggl[ \sum_{a+l+3\leq N\atop {\rm and}\not=0} 
 \frac{{s'}^{a+l+3}}{a+l+3}
\int d\Omega_1 ~n_1^L \! \! \! \!
\sous{1}{f}_a 
+\ln\left({s'\over s_1}\right)\int d\Omega_1
~n_1^L \! \! \! \! \sous{1}{f}_{-l-3} \biggr]\nonumber\\
&+&1\leftrightarrow 2 +o({s'}^N) \;.  
\end{eqnarray*}
Since $\varphi_n$ and all its derivatives $\partial_L\varphi_n$ tend
uniformly towards zero in a given compact $K$, clearly so does the
sequence of real numbers $<{\rm Pf}F_{~\bigl|_{\cal D}},\varphi_n>$,
which shows that ${\rm Pf}F_{~\bigl|_{\cal D}}$ is indeed continuous
({\it QED}).

\begin{definition}
The product (``.'') of $F\in {\cal F}$ and of ${\rm Pf}G\in {\cal
F}'$, and the product of two pseudo-functions ${\rm Pf}F$ and ${\rm
Pf}G$, are defined as

\begin{equation}\label{65}
F~\!.~\!{\rm Pf}G\equiv {\rm Pf}F~\!.~\!{\rm Pf}G\equiv {\rm
Pf}(FG)~\in ~{\cal F}'\;.
\end{equation}
In particular $F~\!.~\!{\rm Pf}G=G~\!.~\!{\rm Pf}F$.
\end{definition}
In the following, we will remove the dot indicating the product and
write indifferently

\begin{equation}\label{66}
F~{\rm Pf}G=G~{\rm Pf}F={\rm Pf}(FG)={\rm Pf}F~{\rm Pf}G=FG~{\rm
Pf}~\!1\;.
\end{equation}
Notice that from the symmetry of the duality bracket we have, $\forall
H\in {\cal F}$,

\begin{equation}\label{67}
<G~{\rm Pf}F,H>={\rm Pf}\int d^3{\bf x}~FGH=<{\rm Pf}F,GH>\;.
\end{equation}
Therefore, when applied to the restriction of pseudo-functions to
${\cal D}$, the product of Definition 6 agrees with the product of a
distribution and a function $\psi\in C^\infty ({\mathbb R}^3$), i.e.

\begin{equation}\label{68}
\forall \varphi\in {\cal D}\;,\quad <\psi~{\rm Pf}F_{~\bigl|_{\cal D}},
\varphi>=<{\rm Pf}F_{~\bigl|_{\cal D}},\psi\varphi>\;.
\end{equation}

\subsection{A Dirac delta-pseudo-function}

Consider, for $\varepsilon\in {\mathbb R}^{+*}$, the Riesz
delta-function ${}_\varepsilon\delta_1$ that we introduced in
(\ref{27}). Since ${}_\varepsilon\delta_1\in {\cal F}$ we can
associate to it the pseudo-function ${\rm
Pf}{}_\varepsilon\delta_1$. Now, Lemma 2 [see (\ref{28})] can be
re-stated by means of the duality bracket as

\begin{equation}\label{69}
\lim_{\varepsilon\to 0}~<{\rm Pf}{}_\varepsilon\delta_1,F>=(F)_1 \;.
\end{equation}
This motivates the following definition.

\begin{definition}
We define the pseudo-function ${\rm Pf}\delta_1$ by

\begin{equation}\label{70}
\forall F\in {\cal F},\quad
<{\rm Pf}\delta_1,F>=(F)_1 \;.
\end{equation}
We then extend the definition of the set ${\cal F}'$ to include this
pseudo-function: ${\rm Pf}\delta_1\in {\cal F}'$.
\end{definition}
Obviously ${\rm Pf}\delta_1$ can be viewed as the ``limit'' [but we
have not defined a topology on ${\cal F}$] of the pseudo-functions
${\rm Pf}{}_\varepsilon\delta_1$ when $\varepsilon\to 0$.  The
restriction of ${\rm Pf}\delta_1$ to ${\cal D}$ is identical to the
usual Dirac measure,

\begin{equation}\label{71}
{\rm Pf}{\delta_1}_{~\bigl|_{\cal D}}=\delta_1\equiv\delta({\bf
x}-{\bf y}_1)\;,
\end{equation}
so that the pseudo-function ${\rm Pf}\delta_1$ appears as a natural
generalization of the Dirac measure in the context of Hadamard parties
finies. In the following, we shall do as if $\delta_1$ would belong to
the original class of functions ${\cal F}$, writing for instance

\begin{equation}\label{72}
<{\rm Pf}F,\delta_1>\equiv <{\rm Pf}\delta_1,F>=(F)_1\;.
\end{equation}
Of course, this equation constitutes in fact the definition of the
bracket $<{\rm Pf}F,\delta_1>$.

\begin{definition} 
For any $F\in{\cal F}$ the pseudo-function ${\rm Pf}(F\delta_1)$ is
defined, consistently with the product ${\rm (\ref{65})}$, by

\begin{equation}\label{73}
\forall G\in {\cal F}\;,\quad
<{\rm Pf}(F\delta_1),G>=(FG)_1 \;.
\end{equation}
We include into ${\cal F}'$ all the pseudo-functions of this type:
${\rm Pf}(F\delta_1)\in {\cal F}'$ (that is, we consider ${\cal
F}'_{\rm new}={\cal F}'+{\cal F}\delta_1+{\cal F}\delta_2$; and we
henceforth drop the ``new").
\end{definition}
Notice that an immediate consequence of the ``non-distributivity'' of
the Hadamard partie finie, namely $(FG)_1\not=(F)_1(G)_1$, is the fact
that

\begin{equation}\label{74}
{\rm Pf}(F\delta_1)\not= (F)_1~{\rm Pf}\delta_1 \;.
\end{equation}
As an example, we have $(r_1)_1=0$; but ${\rm Pf}(r_1\delta_1)$ is not
zero, since $<\!{\rm Pf}(r_1\delta_1),1/r_1\!>=1$ for instance. The
pseudo-function ${\rm Pf}(F\delta_1)$ represents the product of a
delta-function with a function that is singular on its own support,
whereas this product is ill-defined in the standard distribution
theory. However, this object, as seen as a distribution, i.e. when
restricted to the class ${\cal D}$ of smooth functions with compact
support, does exist in the standard theory. Using the Taylor expansion
when $r_1\to 0$ of any $\varphi\in {\cal D}$, that is $\sum_{l\geq
0}{1\over l!}r_1^l n_1^L \partial_L\varphi ({\bf y}_1)$, we obtain

\begin{equation}\label{75}
<{\rm Pf}(F\delta_1)_{\bigl|_{\cal
D}},~\varphi>=(F\varphi)_1=\sum_{l\geq 0}~{1\over l!}
~\partial_L\varphi ({\bf y}_1) \int {d\Omega_1\over 4\pi}~n_1^L
\!\!\sous{1}{f}_{-l}\;,
\end{equation}
where ${}_1f_{-l}$ denotes the coefficient of $1/r_1^l$ in the
expansion of $F$ when $r_1\to 0$. Notice that the sum in (\ref{75}) is
always finite because $l\leq -a_0$, where $a_0=a_0(F)$ is the smallest
exponent of $r_1$ in the expansion of $F$ (see Definition 1).  From
(\ref{75}) we derive immediately the ``intrinsic'' form of the
distribution ${\rm Pf}(F\delta_1)_{\bigl|_{\cal D}}$, that is

\begin{equation}\label{76}
{\rm Pf}(F\delta_1)_{\bigl|_{\cal D}}=\sum_{l\geq 0}~{(-)^l\over l!}
~\partial_L\delta_1 \int {d\Omega_1\over 4\pi}~n_1^L
\!\!\sous{1}{f}_{-l} =\sum_{l\geq 0}~{(-)^l\over l!} ~\Bigl(r_1^l
n_1^L~F\Bigr)_1~\partial_L\delta_1 \;,
\end{equation}
where $\partial_L\delta_1$ denotes the $l$th partial derivative of the
Dirac measure (and where the sums are finite). We have for instance

\begin{equation}\label{77}
{\rm Pf}\Biggl({\delta_1\over r_1^2}\Biggr)_{\bigl|_{\cal
D}}=\frac{1}{6}\Delta\delta_1 \;.
\end{equation}

Note also that the distribution ${\rm Pf}(F\delta_1)_{\bigl|_{\cal
D}}$ can be recovered, quite naturally, from the Laplacian (in the
ordinary distributional sense) of the bracket corresponding to the
``Poisson'' integral of ${\rm Pf}(F\delta_1)$, i.e. formed by ${\rm
Pf}(F\delta_1)$ acting on the function ${\bf x}\rightarrow 1/|{\bf
x}-{\bf x}'|$. For any given ${\bf x}'$, this function belongs to
${\cal F}_{\rm loc}$ and we are still allowed to consider such a
bracket (see also Section V). Thus we define

\begin{equation}\label{78}
G({\bf x}')=-{1\over 4\pi} <{\rm Pf}(F\delta_1),\frac{1}{|{\bf x}-{\bf
x}'|}>=-{1\over 4\pi} \left(\frac{F({\bf x})}{|{\bf x}-{\bf
x}'|}\right)_1\;.
\end{equation}
For ${\bf x}'$ different from the singularity ${\bf y}_1$, we find,
using the Taylor expansion of $1/|{\bf x}-{\bf x}'|$ around ${\bf
y}_1$,

\begin{equation}\label{79}
G({\bf x}')=-{1\over 4\pi} \sum_{l\geq 0}~{(-)^l\over l!} ~\Bigl(r_1^l
n_1^L~F\Bigr)_1~\partial'_L\left({1\over r'_1}\right) \;.
\end{equation}
Clearly the function $G$, if considered as a function of the variable
${\bf x}'$, belongs to ${\cal F}$.  Now, we see from (\ref{76}) that
the ``ordinary'' Laplacian of $G({\bf x}')$ is precisely equal to
${\rm Pf}(F\delta_1)_{\bigl|_{\cal D}}$, namely

\begin{equation}\label{81}
{\Delta' G'}_{\bigl|_{\cal D}}=\sum_{l\geq 0}~{(-)^l\over l!}
~\Bigl(r_1^l n_1^L~F\Bigr)_1~\partial'_L\delta'_1 ={\rm
Pf}(F'\delta'_1)_{\bigl|_{\cal D}}\;.
\end{equation}
Let us point out that $G$ has no partie finie at the point 1:
$(G)_1=0$; so, in order to compute its partie finie at 1, we are not
allowed to replace formally ${\bf x}'$ by ${\bf y}_1$ inside the
defining expression (\ref{78}):

\begin{equation}\label{80}
0=-4\pi (G)_1\not=<{\rm Pf}(F\delta_1),\frac{1}{r_1}>=\Biggl({F\over
r_1}\Biggr)_1=\!\!\sous{1}{{\hat f}}_1\;.
\end{equation}
[The function $G({\bf x}')$ is not continuous at 1, as we can easily
see from its singular expansion (\ref{79}).]

Finally let us mention how to give a sense to a pseudo-function that
would be associated with the square of the
delta-function. $\forall\varepsilon >0$, we have
${}_\varepsilon\delta_1^2\in {\cal F}$, and hence, we can consider the
partie-finie integral of ${}_\varepsilon\delta_1^2 F$. In the limit
$\varepsilon\to 0$ we get

\begin{equation}\label{82}
\lim_{\varepsilon\to 0}<{\rm Pf}{}_\varepsilon\delta_1^2,F>
=\lim_{\varepsilon\to 0}
~{\rm Pf}\int d^3{\bf x}~{}_\varepsilon\delta_1^2~F=0 \;,
\end{equation}
essentially because we have a square $\varepsilon^2$ in factor which
kills any divergencies arising from the integral. Therefore ${\rm
Pf}\delta_1^2$ is (defined to be) identically zero. More generally,

\begin{equation}\label{83}
\forall F\in{\cal F}\;,\quad {\rm Pf}(F\delta_1^2)=0\;,
\end{equation}
and we shall not hesitate to write such identities as

\begin{equation}\label{84}
<{\rm Pf}\delta_1,F\delta_1>=<\delta_1,{\rm Pf}(F\delta_1)>
=<{\rm Pf}(F\delta_1^2),1>=0\;.
\end{equation}
Note also that

\begin{equation}\label{85}
{\rm Pf}(F\delta_1\delta_2)=0\;.
\end{equation}

\section{Derivative of pseudo-functions}

\subsection{A derivative operator on ${\cal F}$}

From now on we shall generally suppose, in order to simplify the
presentation, that the powers of $r_1$ and $r_2$ in the expansions of
$F\in {\cal F}$ around the two singularities are positive or negative
integers ($\in {\mathbb Z}$).  Our aim is to define an appropriate
partial derivative operator acting on the pseudo-functions of the type
${\rm Pf}F$. First of all, we know (Lemma 3) that the restriction of
${\rm Pf}F$ to ${\cal D}$ is a distribution in the ordinary sense, so
we already have at our disposal the derivative operator of
distribution theory \cite{Schwartz}, which is uniquely determined ---
as well as any higher-order derivatives --- by the requirement:
 
\begin{equation}\label{86}
\forall \varphi\in{\cal D}\;,\qquad <\partial_i\bigl({\rm Pf}
F_{~\bigl|_{\cal D}}\bigr)
,\varphi>=-<{\rm Pf}F_{~\bigl|_{\cal D}},\partial_i\varphi>\;.
\end{equation}
It is clear from viewing ${\rm Pf}F_{~\bigl|_{\cal D}}$ as an integral
operator acting on $\varphi$, that (\ref{86}) corresponds to a rule of
``integration by part'' in which the ``all-integrated'' (surface) term
vanishes. In particular the ``integral of a gradient'' is zero.  This
motivates the following definition.

\begin{definition} 
A partial derivative operator $\partial_i$ acting on pseudo-functions
of ${\cal F}'$ is said to satisfy the rule of integration by parts iff

\begin{equation}\label{87}
\forall F, G\in{\cal F}\;,\qquad <\partial_i({\rm Pf}F),G>=
-<\partial_i({\rm Pf}G),F>\;.
\end{equation}
\end{definition}
Notice the symmetry between the two slots of the duality bracket in
(\ref{87}). As an immediate consequence, for a derivative operator
satisfying this rule, we have

\begin{equation}\label{88}
\forall F\in{\cal F}\;,\qquad <\partial_i({\rm Pf}F),F>=0\;.
\end{equation} 
Furthermore, if we assume $\partial_i({\rm Pf}1)=0$ in addition to
Definition 9, then

\begin{equation}\label{89}
\forall F\in{\cal F}\;,\qquad <\partial_i({\rm Pf}F),1>=0\;.
\end{equation}
Of course, both (\ref{88}) and (\ref{89}) correspond to the intuitive
idea that the integral of a gradient (in a ``distributional-extended''
sense) should be zero.

\begin{proposition}
The most general derivative operator on ${\cal F}'$ satisfying the
rule of integration by parts ${\rm (\ref{87})}$ reads

\begin{equation}\label{90}
\partial_i({\rm Pf}F)={\rm Pf}(\partial_iF)+{\sc D}_i[F]~~\in {\cal F}'\;,
\end{equation}
where ${\rm Pf}(\partial_iF)$ represents the ``ordinary'' derivative,
and where the ``distributional'' term ${\sc D}_i[F]={\sc H}_i[F]+{\sc
D}_i^{\rm part}[F]$ is the sum of the general solution of the
homogeneous equation, i.e. a linear functional ${\sc H}_i[F]$ such
that

\begin{equation}\label{92}
\forall F, G\in{\cal F}\;,\qquad
<{\sc H}_i[F],G>+<{\sc H}_i[G],F>=0\;,
\end{equation}
and of the particular solution defined by

\begin{equation}\label{91}
{\sc D}_i^{\rm part}[F]=4\pi ~\!{\rm Pf}\Biggl(n_1^i
\biggl[\case{1}{2}~\!r_1\!\!\!\sous{1}{f}_{-1}+\sum_{k\geq 0}{1\over r_1^k}
\!\!\!\sous{1}{f}_{-2-k}\biggr]\delta_1+1\leftrightarrow 2\Biggr)\;.
\end{equation}
\end{proposition}
When applied on any $G\in{\cal F}$, the particular solution reads

\begin{equation}\label{94}
<{\sc D}_i^{\rm part}[F],G>=\int d\Omega_1 ~n_1^i\biggl[\case{1}{2}\!\!\!
\sous{1}{f}_{-1}\!\!\!\sous{1}{g}_{-1}+\sum_{k\geq 0}\!\!\!\sous{1}{f}_
{-2-k}\!\!\!\sous{1}{g}_k\biggr]
+1\leftrightarrow 2 \;.
\end{equation}

\bigskip\noindent
{\it Proof.} We replace the form (\ref{90}) of the derivative operator
into the rule (\ref{87}) and find $$ <{\sc D}_i[F],G>+<{\sc D}_i[G],F>
=-<{\rm Pf}(\partial_iF),G>-<{\rm Pf}(\partial_iG),F>\;.  $$ The
right-hand side can be readily re-written as the partie-finie integral
of a gradient,

\begin{equation}\label{93'}
<{\sc D}_i[F],G>+<{\sc D}_i[G],F> 
=-{\rm Pf}\int d^3{\bf x}~\partial_i(FG)\;.
\end{equation}
Now we know from (\ref{23}) that the integral of a gradient is equal
to the partie finie of the surface integrals around the singularities
when the surface areas shrink to zero; thus $$ <{\sc D}_i[F],G>+<{\sc
D}_i[G],F> =4\pi (n_1^i r_1^2 FG)_1+1\leftrightarrow 2\;.  $$ We
replace into the right side $F$ and $G$ by their expansions around 1,
and after an easy calculation we arrive at

\begin{equation}\label{93''} 
<{\sc D}_i[F],G>+<{\sc D}_i[G],F> =\int
d\Omega_1~n_1^i\Biggl[\!\!\!\sous{1}{f}_{-1}\!\!\!\sous{1}{g}_{-1}+\sum_{k\geq
0}\biggl(\!\!\!\sous{1}{f}_{-2-k}\!\!\!\sous{1}{g}_k+\!\!\!\sous{1}
{f}_k\!\!\!\sous{1}{g}_{-2-k}\biggr)\Biggr]
+1\leftrightarrow 2\;.\nonumber\\
\end{equation}
It is clear that the particular solution given by (\ref{91}) or
(\ref{94}) solves the latter equation. As a consequence, the most
general solution is simply obtained by adding the general solution of
the homogeneous equation, i.e. (\ref{93''}) with zero in the right
side, which is precisely a ${\sc H}_i[F]$ satisfying the
``anti-symmetry'' property $<{\sc H}_i[F],G>+<{\sc H}_i[G],F>=0$.
{\it QED}.

As we see from Proposition 3, the rule of integration by parts does
not permit, unlike in the case of distribution theory [see
(\ref{86})], to fully specify the derivative operator. Obviously, we
must supplement the rule by another statement indicating the cases for
which the new derivative should reduce to the ``ordinary'' one,
i.e. when we should have $\partial_i({\rm Pf}F)={\rm
Pf}(\partial_iF)$. Clearly, we would like to recover the ordinary
derivative in the cases where the function is ``not too much
singular''. In the following, we shall require essentially that our
derivative reduces to the ordinary one when the function $F$ is {\it
bounded} near the singularities [in addition of belonging to
$C^\infty({\mathbb R}^3-\{{\bf y}_{1,2}\})$], in the sense that there
exists a neighbourhood ${\cal N}$ containing the two singularities
${\bf y}_1$ and ${\bf y}_2$ and a constant $M\in {\mathbb R}^{+*}$
such that ${\bf x}\in {\cal N}~\Rightarrow |F({\bf x})|\leq M$. Let us
refer to the coefficients of the negative powers of $r_1$ and $r_2$ in
the expansions of $F$, i.e. the ${}_1f_{-1-k}$'s and ${}_2f_{-1-k}$'s
where $k\in {\mathbb N}$, as the {\it singular} coefficients of $F$
(recall that we assumed that the powers of $r_1$ and $r_2$ are
integers). Clearly, a function is bounded near the singularities if
and only if all its singular coefficients vanish. This means that we
shall require that the distributional term ${\sc D}_i[F]$, which is a
linear functional of the coefficients in the expansions of $F$, should
depend only on the singular coefficients ${}_1f_{-1-k}$ and
${}_2f_{-1-k}$ of $F$. This is already the case of our particular
solution ${\sc D}_i^{\rm part}[F]$ in (\ref{91}). We now look for the
most general possible ${\sc H}_i[F]$ depending on the ${}_1f_{-1-k}$'s
(and $1\leftrightarrow 2$).

All the singular coefficients admit some spherical-harmonics or
equivalently STF expansions of the type (\ref{10})-(\ref{11}), with
STF-tensorial coefficients ${}_1{\hat f}_{-1-k}^L$ [where $L=i_1\cdots i_l$;
see (\ref{10}) for definition], so we are led to requiring that ${\sc
H}_i[F]$ be the most general (linear) functional of the STF tensors
${}_1{\hat f}_{-1-k}^L$ and $1\leftrightarrow 2$. Moreover, we demand that
${\sc H}_i[F]$, like ${\sc D}_i^{\rm part}[F]$, is proportional to the
Dirac pseudo-function ${\rm Pf}\delta_1$ (as we shall see, the
gradient of ${\rm Pf}\delta_1$ is itself proportional to ${\rm
Pf}\delta_1$ so there is no loss of generality). Now, we have also to
take into account the fact that the dimensionality of ${\sc H}_i[F]$
should be compatible with the one of ${\rm Pf}(\partial_iF)$. Endowing
${\mathbb R}^3$ with a unit of length to measure the space
coordinates, the Dirac pseudo-function ${\rm Pf}\delta_1$ takes the
dimension of the inverse cube of a length, and ${\sc H}_i[F]$ the
dimension of $F$ divided by this length (in physical applications, we
do not want to introduce any special physical scale). We conclude that
${\sc H}_i[F]$ must be of the general form:

\begin{equation}\label{94'}
{\sc H}_i[F]=\sum_{k\geq 0}\!~\sum_{l=0}^{+\infty}{\rm
Pf}\biggl(\Bigl[\alpha_{k,l}~\!{\hat n}_1^{iL}\!\!\!\sous{1}{{\hat
f}}_{-1-k}^{~L}+\beta_{k,l}~\!n_1^{L}\!\!\!\sous{1}{{\hat
f}}_{-1-k}^{~iL}\Bigr] r_1^{1-k}\delta_1\biggr)+1\leftrightarrow 2\;,
\end{equation}
where the $\alpha_{k,l}$'s and $\beta_{k,l}$'s denote some purely
constant numerical coefficients (and where, as usual, the sum over $k$
is finite). Applying this ${\sc H}_i[F]$ on any $G$ we readily obtain

\begin{equation}\label{94''}
<{\sc H}_i[F],G>=\sum_{k\geq 0}\!~\sum_{l=0}^{+\infty}\case{l!}{
(2l+1)!!}\biggl[\case{l+1}{2l+3}\alpha_{k,l}\!\!\!\sous{1}{{\hat
f}}_{-1-k}^{~L}\!\!\!\sous{1}{{\hat
g}}_{-1+k}^{~iL}+\beta_{k,l}\!\!\!\sous{1}{{\hat
f}}_{-1-k}^{~iL}\!\!\!\sous{1}{{\hat
g}}_{-1+k}^{~L}\biggr]+1\leftrightarrow 2\;.
\end{equation}
At last we must impose the anti-symmetry condition (\ref{92}). For any
$G$ whose all singular coefficients vanish we have $<{\sc
H}_i[G],F>=0$; then, the anti-symmetry condition tells us that
(\ref{94''}) should be identically zero for any such $G$ and any
$F$. Therefore, we must have $\alpha_{k,l}=0$ and $\beta_{k,l}=0$
whenever $k\geq 1$, so we are left with only the coefficients
$\alpha_{0,l}$ and $\beta_{0,l}$, and the condition (\ref{92}) now
implies 
$$ 0=\sum_{l=0}^{+\infty}\case{l!}{
(2l+1)!!}\Bigl(\case{l+1}{2l+3}\alpha_{0,l}+\beta_{0,l}
\Bigr)\biggl[\!\!\!\sous{1}{{\hat
f}}_{-1}^{~L}\!\!\!\sous{1}{{\hat g}}_{-1}^{~iL}+\!\!\!\sous{1}{{\hat
f}}_{-1}^{~iL}\!\!\!\sous{1}{{\hat
g}}_{-1}^{~L}\biggr]+1\leftrightarrow 2\;, $$ which can clearly be
satisfied only if (and only if)
$\case{l+1}{2l+3}\alpha_{0,l}+\beta_{0,l}=0$. Thus, posing
$\alpha_l\equiv \alpha_{0,l}$, we have just proved:

\begin{lemma} 
The most general ${\sc H}_i[F]$ that vanishes for any bounded function
$F\in {\cal F}$ and possesses the correct dimension depends only on
(the STF-harmonics of) the singular coefficients ${}_1f_{-1}$ and
${}_2f_{-1}$ and is given by

\begin{equation}\label{95}
{\sc H}_i[F]=\sum_{l=0}^{+\infty}~\!\alpha_l~\!{\rm
Pf}\biggl(\Bigl[{\hat n}_1^{iL}\!\!\!\sous{1}{{\hat
f}}_{-1}^{~L}-\case{l+1}{2l+3}~\!n_1^{L}\!\!\!\sous{1}{{\hat
f}}_{-1}^{~iL}\Bigr] r_1\delta_1\biggr)+1\leftrightarrow 2\;,
\end{equation}
where the $\alpha_l$'s form a countable set of arbitrary numerical
coefficients.
\end{lemma}
[The angular dependence of the first term in (\ref{95}) is expressed
by means of the STF tensor ${\hat n}_1^{iL}$.]  Equivalently we have

\begin{equation}\label{95'}
<{\sc H}_i[F],G>=\sum_{l=0}^{+\infty}~\!\alpha_l~\!\case{(l+1)!}{
(2l+3)!!}\biggl[\!\!\!\sous{1}{{\hat f}}_{-1}^{~L}\!\!\!\sous{1}{{\hat
g}}_{-1}^{~iL}-\!\!\!\sous{1}{{\hat f}}_{-1}^{~iL}\!\!\!\sous{1}{{\hat
g}}_{-1}^{~L}\biggr]+1\leftrightarrow 2\;.
\end{equation}
This expression is anti-symmetric in the exchange $F\leftrightarrow G$
as required.

To sum up, we have obtained the most general derivative operator
$\partial_i({\rm Pf}F)={\rm Pf}(\partial_iF)+{\sc D}_i[F]$ that
satisfies the rule of integration by parts and depends only on the
singular coefficients of $F$. The distributional term ${\sc D}_i[F]$
is the sum of a ``particular'' solution fully specified by (\ref{91})
or (\ref{94}), and of a ``homogeneous'' solution given by (\ref{95})
or (\ref{95'}) in terms of an infinite set of arbitrary numerical
coefficients $\alpha_l\in {\mathbb R}$ (and $l\in {\mathbb N}$). In
Section VIII we shall see how one can reduce the arbitrariness of the
definition of the derivative to only one single coefficient $K\in
{\mathbb R}$.

\subsection{Some properties of the derivative}

At this stage, one can already investigate some properties of the
distributional term ${\sc D}_i[F]={\sc D}_i^{\rm part}[F]+{\sc
H}_i[F]$, using the fact that the yet un-specified $<{\sc H}_i[F],G>$
depends only on ${}_1f_{-1}$ and ${}_1g_{-1}$ (and $1\leftrightarrow
2$).  Let us first check that the derivative operator, when restricted
to the smooth and compact-support functions of ${\cal D}$, reduces to
the distributional derivative of distribution theory
\cite{Schwartz}. This must actually be true since the fundamental
property (\ref{86}) of the distributional derivative is a particular
case of our rule of integration by parts, and because the derivative
of $\varphi\in {\cal D}$ reduces to the ordinary one. However, it is
instructive to verify directly this fact using the expression
(\ref{91}). Applying ${\sc D}_i[F]$ on $\varphi\in {\cal D}$ and using
the Taylor expansion of $\varphi$ around 1: $\varphi=\sum_{k\geq
0}\case{1}{k!}r_1^k n_1^K (\partial_K\varphi)({\bf y}_1)$, we obtain

\begin{eqnarray*}
<{\sc D}_i[F],\varphi>&=&\sum_{k\geq 0}{1\over k!}
(\partial_K\varphi)({\bf y}_1) \int
d\Omega_1~n_1^in_1^K\!\!\!\sous{1}{f}_{-2-k} +1\leftrightarrow 2\;.
\end{eqnarray*}
Hence the intrinsic expression of the distributional terms on ${\cal
D}$,

\begin{equation}\label{96}
{\sc D}_i[F]_{\!~\bigl|_{\cal D}} = \sum_{k\geq 0}{(-)^k\over k!}
\partial_K\delta_1 \int d\Omega_1~n_1^in_1^K\!\!\!\sous{1}{f}_{-2-k}
+1\leftrightarrow 2\;,
\end{equation}
which agrees with the distributional part of the derivative of a
function with tempered singularities in distribution theory. For
example, we can write

\begin{equation}\label{96'}
{\sc D}_i\biggl[{1\over r_1^3}\biggr]_{\!~\bigl|_{\cal D}} =
-{4\pi\over 3} ~\!\partial_i\delta_1\;.
\end{equation}

However, when acting on functions of the full set ${\cal F}$, the
derivative generally leads to properties which have no equivalent in
distributional theory. For instance, although the distributional
derivative of $1/r_1^2$ reduces on ${\cal D}$ to the ordinary
derivative, i.e. ${\sc D}_i[\case{1}{r_1^2}]_{\!~\bigl|_{\cal D}}=0$,
on ${\cal F}$ it does not:

\begin{equation}\label{97}
\partial_i\biggl({\rm Pf}{1\over r_1^2}\biggr)={\rm Pf}\biggl(-2{n_1^i\over r_1^3}
+4\pi \!~n_1^i\delta_1\biggr)\;.
\end{equation}
For the distributional derivative of $1/r_1^3$ on ${\cal F}$ we find

\begin{equation}\label{98}
\partial_i\biggl({\rm Pf}{1\over r_1^3}\biggr)={\rm Pf}\biggl(-3{n_1^i\over r_1^4}
+4\pi {n_1^i\over r_1}\delta_1\biggr)\;.
\end{equation}
The expression of the distributional term is apparently different from
the corresponding result (\ref{96'}) in distribution theory. However
we shall see after learning how to differentiate the Dirac
pseudo-function ${\rm Pf}\delta_1$ that the distributional term ${\sc
D}_i[\case{1}{r_1^3}]$ takes in fact the same form on ${\cal F}$ as on
${\cal D}$ [see (\ref{109}) below].

We come now to an important point. In this paper we have defined a
``pointwise'' product of pseudo-functions (see Definition 6), which
reduces to the ordinary product in all the cases where the functions
are regular enough. For instance, it coincides with the ordinary
product for $C^\infty$ functions, or even continuous or locally
integrable functions (adopting the class ${\cal F}_{\rm loc}$). Next,
we introduced a derivative operator that acts merely as the ordinary
derivative for a large class of not-too-singular functions (those
which are bounded near the singularities, see Proposition 3). In
particular, the derivative is equal to the ordinary one when the
functions are $C^1$ at the location of the two singularities.
However, we know from a theorem of Schwartz \cite{Schwartz54} that it
is impossible to define a multiplication for distributions having the
previous properties and such that the distributional derivation
satisfies the standard formula for the derivation of a product
(Leibniz's rule). In agreement with that theorem, we find that the
derivative operator defined by (\ref{90})-(\ref{91}) does not obey in
general the Leibniz rule, whereas it does satisfy it by definition in
an ``integrated sense'', namely:

\begin{equation}\label{99}
<\partial_i[{\rm Pf}(FG)],1>=0=<\partial_i({\rm
Pf}F)G+F\partial_i({\rm Pf}G),1>\;.
\end{equation}
However it does not satisfy the Leibniz rule in a ``local sense'',
i.e. we have, generically for two functions $F,G\in {\cal F}$,

\begin{equation}\label{100}
\partial_i[{\rm Pf}(FG)]-\partial_i({\rm Pf}F)G-F\partial_i({\rm Pf}G)\not=0\;.
\end{equation}
This means that, {\it a priori},

\begin{equation}\label{101}
<\partial_i[{\rm Pf}(FG)],H>-<\partial_i({\rm Pf}F),GH>-<\partial_i(
{\rm Pf}G),FH>\not=0\;,
\end{equation}
or, equivalently, since the Leibniz rule is satisfied by the ordinary
derivative,

\begin{equation}\label{102}
<{\sc D}_i[FG],H>-<{\sc D}_i[F],GH>-<{\sc D}_i[G],FH>\not=0\;.
\end{equation}
Actually, in accordance with the theorem in \cite{Schwartz54},
(\ref{100}) must be true even when the pseudo-function is regarded as
a distribution on ${\cal D}$. To check this, let us compute the left
side of (\ref{102}) in the case where ${\sc D}_i$ is the particular
solution ${\sc D}_i^{\rm part}$ defined by (\ref{91}), and where $H$
is equal to some $\varphi\in {\cal D}$. We employ the Taylor expansion
of $\varphi$ around 1 and 2, and, strictly following the definition of
the distributional term in (\ref{91}), we arrive at

\begin{eqnarray}\label{103}
&&\qquad\qquad\biggl[{\sc D}_i^{\rm part}[FG]-F {\sc D}_i^{\rm
part}[G]-G{\sc D}_i^{\rm part}[F]\biggr]_{\bigl|_{\cal D}}\nonumber\\
&&=\sum_{k\geq 1}{(-)^k\over k!} \partial_K\delta_1 \int
d\Omega_1~n_1^in_1^K
\biggl[\case{1}{2}\!\!\!\sous{1}{f}_{-1}\!\!\!\sous{1}{g}_{-1-k}
+\case{1}{2}\!\!\!\sous{1}{f}_{-1-k}\!\!\!\sous{1}{g}_{-1}-\sum_{j
=0}^{k}\!\!\!\sous{1}{f}_{-1-j}\!\!\!\sous{1}{g}_{j-k-1}\biggr]
+1\leftrightarrow
2\;.
\end{eqnarray}
The right side of (\ref{103}) equals ${2\pi\over 3}\partial_i\delta_1$
in the case where $F={1\over r_1}$ and $G={1\over r_1^2}$ for
instance.  It is not possible to add a homogeneous solution of the
form (\ref{95}) so as to get always zero.  As the result (\ref{103})
depends only on the singular coefficients of $F$ and $G$, we recover
the Leibniz rule whenever $F$ or $G$ is bounded near the
singularities. Besides, we can verify directly on (\ref{103}) that the
Leibniz rule is indeed true in an integrated sense, since the integral
over ${\mathbb R}^3$ of (\ref{103}) picks up only the term with $k=0$
which gives no contribution.

\subsection{Derivative of the Dirac pseudo-function}

In this subsection we compute the distributional term $<{\sc
D}_i[F],G>$ given by the sum of (\ref{94}) and (\ref{95'}) assuming
that either $F$ or $G$ is equal to the Riesz delta-function
${}_\varepsilon\delta_1=\case{\varepsilon(\varepsilon-1)}
{4\pi}r_1^{\varepsilon-3}$
for some small $\varepsilon>0$. (We come back for a moment to
Definition 1 in which the powers of $r_1$ and $r_2$ in the expansions
of $F$ or $G$ are real.) We notice first that the terms depending on
the singular coefficients ${}_1f_{-1}$ and ${}_1g_{-1}$ are present
only when the exponent $-1$ belongs to both families of indices
$(a_i)_{i\in {\mathbb N}}$ corresponding to $F$ and $G$ (remind
Definition 1). This means that, choosing $\varepsilon$ to be different
from 2, these terms will not contribute to the present calculation,
and in particular that the homogeneous part $<{\sc H}_i[F],G>$ will
always give zero, provided that either $F$ or $G$ is equal to
${}_\varepsilon\delta_1$. From the expression (\ref{94}) we get

\begin{mathletters}\label{106'}\begin{eqnarray}
<{\sc D}_i[{}_\varepsilon\delta_1],G>&=&\varepsilon(1-\varepsilon)\int
{d\Omega_1\over
4\pi}~n_1^i\!\!\!\sous{1}{g}_{1-\varepsilon}\;,\label{106'a}\\ <{\sc
D}_i[F],{}_\varepsilon\delta_1>&=&\varepsilon(1-\varepsilon)
\sum_{l \ge 0} \frac{(-)^l}{l!}
\!\!\sous{1}{\partial}_{~\!L} r_{12}^{\varepsilon-3} 
\int {d\Omega_2\over 4\pi}~n_2^{iL} \!\!\!\sous{2}{f}_{-2-l}\;.\label{106'b}
\end{eqnarray}\end{mathletters}$\!\!$
Furthermore, by choosing $\varepsilon$ smaller than the spacing
between some exponents $a_i$ of $G$ (specifically $\varepsilon
<1-a_{i_1}$ with $a_{i_1}$ is such that $a_{i_1}<1\leq a_{i_1+1}$) we
can arrange for having ${}_1g_{1-\varepsilon}=0$ so that (\ref{106'a})
becomes identically zero. Anyway, in the limit $\varepsilon\to 0$ we
come up formally with both relations $<{\sc D}_i[\delta_1],G>=0$ and $<{\sc
D}_i[F],\delta_1>=0$. The former tells us that the distributional
derivative of ${\rm Pf}\delta_1$ reduces to the ordinary one, i.e.

\begin{equation}\label{106''}
\partial_i({\rm Pf}\delta_1)={\rm Pf}(\partial_i\delta_1)\;.
\end{equation}
The latter [that we already knew from (\ref{84})] shows via the rule
of integration by parts that the action of $\partial_i({\rm
Pf}\delta_1)$ over any function $F\in {\cal F}$ is equal to minus the
action of ${\rm Pf}\delta_1$ over the derivative $\partial_iF$.

\begin{definition}
The derivative of the Dirac pseudo-function ${\rm Pf}\delta_1$ is
defined by

\begin{equation}\label{107}
\forall F\in{\cal F}\;,\qquad <\partial_i({\rm Pf}\delta_1),F>
=-<{\rm Pf}\delta_1,\partial_iF>\equiv-(\partial_iF)_1\;.
\end{equation}
\end{definition}
We can summarize the properties of the derivative of the Dirac
pseudo-function by writing the successive identities

$$ <\partial_i({\rm Pf}\delta_1),F>=<{\rm
Pf}(\partial_i\delta_1),F>=-<{\rm
Pf}\delta_1,\partial_iF>=-(\partial_iF)_1\;, $$ as well as similar
identities obtained by exchanging the roles of $F$ and $\delta_1$, $$
<\partial_i({\rm Pf}F),\delta_1>=<{\rm
Pf}(\partial_iF),\delta_1>=-<{\rm
Pf}F,\partial_i\delta_1>=(\partial_iF)_1\;.  $$

\begin{lemma}
The intrinsic form of the derivative of the Dirac pseudo-function is
\begin{equation}\label{108}
\partial_i({\rm Pf}\delta_1)=-{\rm Pf}\biggl(3{n_1^i\over r_1}\delta_1\biggr)\;.
\end{equation}
\end{lemma}
The proof is evident from using the identity (\ref{16}).  The form
(\ref{108}) [with (\ref{106''})] is quite useful in practice; for
instance, it permits us to re-write the derivative of the
pseudo-function ${\rm Pf}({1\over r_1^3})$ as computed in (\ref{98})
into the form

\begin{equation}\label{109}
\partial_i\biggl({\rm Pf}{1\over r_1^3}\biggr)={\rm Pf}\biggl
(-3{n_1^i\over r_1^4}-{4\pi\over 3} \partial_i\delta_1\biggr)\;,
\end{equation}
where the distributional term takes the same form as in the
distribution theory [compare with (\ref{96'})].

The preceding definition and lemma are easily extended to the case of
the pseudo-functions ${\rm Pf}(F\delta_1)$. The derivative of these
objects is defined by the mean of the relation

\begin{equation}\label{110}
<\partial_i[{\rm Pf}(F\delta_1)],G>=-<{\rm
Pf}(F\delta_1),\partial_iG>=-(F\partial_iG)_1\;.
\end{equation}
Then, from the identity ${\rm (\ref{16})}$, we readily get the
intrinsic form

\begin{equation}\label{111}
\partial_i[{\rm Pf}(F\delta_1)]=
{\rm Pf}\Biggl[r_1^3\partial_i\biggl({F\over
r_1^3}\biggr)\delta_1\Biggr]\;.
\end{equation}
Notice the interesting particular case

\begin{equation}\label{112}
\partial_i[{\rm Pf}(r_1^3\delta_1)]=0\;,
\end{equation}
which is also an immediate consequence of (\ref{17}). Finally, let us
mention that the Leibniz rule happens to hold in the special case
where one of the pseudo-functions is of the type ${\rm
Pf}(G\delta_1)$, i.e.

\begin{equation}\label{112'}
\partial_i[{\rm Pf}F~\!\!.~\!{\rm Pf}(G\delta_1)]=
\partial_i({\rm Pf}F)~\!\!.~\!{\rm Pf}(G\delta_1)+{\rm Pf}F~\!\!.
~\!\partial_i[{\rm Pf}(G\delta_1)]
\end{equation}
(the verification is straightforward).

\section{Multiple derivatives}

\subsection{General construction}

From Proposition 3 we can give a meaning to

\begin{equation}\label{104}
<\partial_i({\rm Pf}F),G> ={\rm Pf}\int d^3{\bf x}~\partial_iFG +<{\sc
D}_i[F],G>\;,
\end{equation}
which will be also denoted $<\partial_i({\rm Pf}F),{\rm Pf}G>$. We now
define the more complicated object $<\partial_i({\rm
Pf}F),\partial_j({\rm Pf}G)>$. Since the distributional term ${\sc
D}_i[F]$ has the form ${\rm Pf}(H\delta_1)$ plus $1\leftrightarrow 2$,
and because (\ref{83})-(\ref{85}) entail such identities as $< {\rm
Pf} (G\delta_1),{\rm Pf}(H\delta_1)>=0=<{\rm Pf}(G\delta_1),{\rm
Pf}(H\delta_2)>$, we deduce that the duality bracket applied on any
two distributional terms is always zero:

\begin{equation}\label{104'}
\forall F, G\in {\cal F}\;,\quad <{\sc D}_i[F],{\sc D}_j[G]>=0\;.
\end{equation}
When constructing the bracket $<\partial_i({\rm Pf}F),\partial_j({\rm
Pf}G)>$ we shall meet a product of two distributional terms which
gives zero by (\ref{104'}), and we shall be left only with the
ordinary part as well as the two cross terms involving one
distributional term. Therefore,

\begin{eqnarray}\label{105}
<\partial_i({\rm Pf}F),\partial_j({\rm Pf}G)> &=&{\rm Pf}\int d^3{\bf
x}~\partial_iF\partial_jG\nonumber\\ &+&<{\sc
D}_i[F],\partial_jG>+<{\sc D}_j[G],\partial_iF>\;.
\end{eqnarray}
[The ordinary part can equivalently be written as $$ {\rm Pf}\int
d^3{\bf x}~\partial_iF\partial_jG =<{\rm Pf}(\partial_iF),{\rm
Pf}(\partial_jG)>=<{\rm
Pf}(\partial_iF),\partial_jG>=<\partial_iF,{\rm Pf}(\partial_jG)>\;. ]
$$ We now intend to introduce the second-order derivative
operator. The generalization to any $l$th-order derivative is
straightforward and will be stated without proof. By extending the
rule of integration by parts presented in Definition 9, we are led,
quite naturally, to require that

\begin{equation}\label{113}
\forall F,G\in {\cal F},\quad
<\partial_{ij}({\rm Pf}F),G>=-<\partial_j({\rm Pf}F),\partial_i({\rm
Pf}G)>\;,
\end{equation}
where the object $<\partial_j({\rm Pf}F),\partial_i({\rm Pf}G)>$ has
just been given in (\ref{105}). For the moment, we are careful at
distinguishing the order of the indices $i$ and $j$. Let us look for
the expression of the distributional term ${\sc D}_{ij}[F]$
corresponding to the double derivative, {\it viz}

\begin{equation}\label{114}
\partial_{ij}({\rm Pf}F)={\rm Pf}(\partial_{ij}F)+{\sc D}_{ij}[F]\;,
\end{equation}
in terms of the single-derivative term ${\sc D}_i[F]$. Inserting
(\ref{114}) into the required property (\ref{113}) we arrive
immediately at $$ <{\sc D}_{ij}[F],G>=-{\rm Pf}\int d^3{\bf
x}~\partial_i\bigl(\partial_jF\!~G\bigr) -<{\sc D}_j[F],\partial_iG>-<{\sc
D}_i[G],\partial_jF>\;.  $$ Next recall the formula (\ref{93'}) which
tells us that any partie-finie integral of a gradient is the sum of
two distributional contributions. Using this property we obtain the
simple result

\begin{eqnarray}\label{115}
<{\sc D}_{ij}[F],G>&=&<{\sc
D}_i[\partial_jF],G>-<{\sc D}_j[F],\partial_iG>\nonumber\\ &=&<{\sc
D}_i[\partial_jF],G>+<\partial_i{\sc D}_j[F],G>\;.
\end{eqnarray}
The formula (\ref{110}) allowed us to obtain the second equality; so
the intrinsic form of the second-order distributional term is obtained
as

\begin{equation}\label{115'}
{\sc D}_{ij}[F]={\sc D}_i[\partial_jF]+\partial_i{\sc D}_j[F]\;.
\end{equation}
This result is easily extendible to any multiple derivatives,
demanding that, to any order $l$,

\begin{equation}\label{116}
<\partial_{i_1i_2\cdots i_l}({\rm Pf}F),G>=-<\partial_{i_2\cdots
i_l}({\rm Pf}F),\partial_{i_1}({\rm Pf}G)>\;,
\end{equation}
where the right side is obtained in a way similar to (\ref{105}). We
can even impose the more general rule of integration by parts, that
for {\it any} $k=1,\cdots,l$,

\begin{equation}\label{119}
<\partial_{i_1i_2\cdots i_l}({\rm
Pf}F),G>=(-)^k<\partial_{i_{k+1}i_{k+2}\cdots i_l}({\rm
Pf}F),\partial_{i_ki_{k-1}\cdots i_1}({\rm Pf}G)>\;.
\end{equation}
Then the following is proved by induction over $l$.

\begin{proposition} If a multi-derivative operator  

\begin{equation}\label{120}
\partial_{i_1i_2\cdots i_l}({\rm Pf}F)={\rm Pf}(\partial_{i_1i_2\cdots i_l}F)
+{\sc D}_{i_1i_2\cdots i_l}[F]\;,
\end{equation}
satisfies the rule of integration by parts ${\rm (\ref{116})}$ or
${\rm (\ref{119})}$, then the $l$-th order distributional term ${\sc
D}_{i_1i_2\cdots i_l}[F]$ is given in terms of the first-order ${\sc
D}_{i_k}[F]$'s by

\begin{equation}\label{120'}
{\sc D}_{i_1i_2\cdots i_l}[F]=\sum_{k=1}^l\partial_{i_1\cdots
i_{k-1}}{\sc D}_{i_k}[\partial_{i_{k+1}\cdots i_l}F]\;.
\end{equation}
\end{proposition}
Recall that this result is valid for any distributional derivative of
the form given by Proposition 3, i.e. ${\sc D}_i[F]={\sc D}_i^{\rm
part}[F]+{\sc H}_i[F]$.  Therefore, the rule of integration by parts
has permitted us to construct uniquely all higher-order derivatives
from a given choice of first-order derivative ${\sc D}_i[F]$,
i.e. from a given choice of ``homogeneous'' solution ${\sc
H}_i[F]$. Notice that {\it a priori} this construction does not yield
some commuting multi-derivatives (i.e. the Schwarz lemma is not valid
in general), because evidently the right side of the formula
(\ref{120'}) is not necessarily symmetric in all its indices. However,
as a central result of this paper, we shall show now that it is
possible to find an initial ${\sc H}_i[F]$ such that the derivatives
do commute to any order.

\begin{theorem} 
The most general derivative operator $\partial_i({\rm Pf}F)={\rm
Pf}(\partial_iF)+{\sc D}_i[F]$ such that

(i) the distributional term ${\sc D}_i[F]$ depends only on the
singular coefficients of $F$,

(ii) all multi-derivatives satisfy the rule of integration by parts,

(iii) all multi-derivatives commute (i.e. the ${\sc D}_{i_1i_2\cdots
i_l}[F]$'s are symmetric in $i_1i_2\cdots i_l$),

\noindent
is given by

\begin{equation}\label{121}
{\sc D}_i[F]=4\pi\sum_{l=0}^{+\infty}~\!{\rm
Pf}\Biggl(~\!C_l~\!\biggl[n_1^{iL}\!\!\!\sous{1}{{\hat
f}}_{-1}^{~L}-n_1^{L}\!\!\!\sous{1}{{\hat f}}_{-1}^{~iL}\biggr]
r_1\delta_1 +\sum_{k\geq 0}{n_1^{iL}\over r_1^k}\!\!\!\sous{1}{{\hat
f}}_{-2-k}^{~L}\delta_1\Biggr)+1\leftrightarrow 2\;,
\end{equation}
where the coefficients $C_l=(l+1)\Bigl[K+\sum_{j=1}^l{1\over
j+1}\Bigr]$ depend on an arbitrary constant $K$.
\end{theorem}
(Actually the theorem states that the derivative operator depends {\it
a priori} on two different constants $K_1$ and $K_2$ for each of the
two singularities. In the following we shall assume for simplicity
that the constants are the same, so that the way to differentiate does
not distinguish between the different singularities.) Notice that
${\sc D}_i[F]$ differs from the particular solution ${\sc D}_i^{\rm
part}[F]$ given by (\ref{91}) 
only in the terms depending on the ``least singular''
coefficients ${}_1f_{-1}$ and ${}_2f_{-1}$.

\bigskip\noindent
{\it Proof.} According to the assumptions (i) and (ii) we already know
(see Proposition 3 and Lemma 4) that the distributional term must be
of the form ${\sc D}_i[F]={\sc D}_i^{\rm part}[F]+{\sc H}_i[F]$, where
the particular solution is given explicitly by (\ref{91}), and where
the homogeneous term takes the form (\ref{95}) depending on a set of
arbitrary coefficients $\alpha_l$. Furthermore, we know from
Proposition 4 that all higher-order derivatives are generated from the
first-order one in the way specified by (\ref{120'}).  It only remains
to show that the coefficients $\alpha_l$ can be {\it computed} in
order that the assumption (iii) of commutation of derivatives be
fulfilled, and that the derivative is given by (\ref{121}).

What we want then is to impose the symmetry of ${\sc D}_{ij}[F]$ in
$ij$. We compute the anti-symmetric projection $[ij]\equiv {ij-ji\over
2}$ of the second-order distributional term associated with the
particular solution (\ref{91}),

\begin{equation}\label{121'}
{\sc D}_{[ij]}^{\rm part}[F]={\sc D}_{[i}^{\rm
part}[\partial_{j]}F]+\partial_{[i}{\sc D}_{j]}^{\rm part}[F]\;.
\end{equation}
The first term is readily obtained from (\ref{13}) which tells us that
the $a$th coefficient in the $r_1$-expansion of the gradient is
${}_1f_a(\partial_jF)=(a+1)n_1^j{}_1f_{a+1}+d_1^{~\!\!j}~\!{}_1f_{a+1}$. On
the other hand, the second term in (\ref{121'}) comes directly from
using the formula (\ref{111}). It follows that the anti-symmetric
projection depends only on the expansion coefficients ${}_1f_0$,
${}_1f_{-1}$ and $1\leftrightarrow 2$ through the simple formula,

\begin{equation}\label{121''}
{\sc D}_{[ij]}^{\rm part}[F]=2\pi ~\! {\rm
Pf}\biggl(n_1^{[i}~\!\Bigl[r_1
~\!d_1^{~\!\!j]}\!\!\!\sous{1}{f}_0+~\!d_1^{~\!\!j]}\!\!\!\sous{1}
{f}_{-1}\Bigr]\delta_1\biggr)+1\leftrightarrow
2\;,
\end{equation}
or, using the relation (\ref{13'}) for the operator $d_1^{~\!\!j}$,

\begin{equation}\label{121'''}
{\sc D}_{[ij]}^{\rm part}[F]=2\pi ~\!\sum_{l=0}^{+\infty} (l+1)
~\!{\rm Pf}~\!\biggl(n_1^{L[i}~\!\Bigl[r_1 \!\!\!\sous{1}{\hat
f}_0^{~j]L}+\!\!\!\sous{1}{f}_{-1}^{~j]L}\Bigr]\delta_1\biggr)+1\leftrightarrow
2\;.
\end{equation}
Note that by applying this on any $G$, we get $$ <{\sc D}_{[ij]}^{\rm
part}[F],G>=-2\pi \sum_{l=0}^{+\infty}\case{\!~(l+1)\!~(l+1)!}{
(2l+3)!!}\biggl(\!\!\sous{1}{{\hat f}}_0^{~L[i}
\!\!\sous{1}{{\hat g}}_{-1}^{~j]L}+\!\!\sous{1}{{\hat f}}_{-1}^{~L[i}
\!\!\sous{1}{{\hat g}}_{~\!0}^{~j]L}\biggr)+1\leftrightarrow 2\;.
$$ Next, we add the homogeneous solution. By performing a computation
similar as the previous one (but a bit more involved) we find, based
on the expression (\ref{95}),

\begin{equation}\label{122}
{\sc H}_{[ij]}[F]=~\!\sum_{l=0}^{+\infty}
\case{l+1}{2l+3}[(l+2)\alpha_l-(l+1)\alpha_{l+1}] ~\!{\rm
Pf}\biggl(n_1^{[i}~\!\Bigl[r_1 \!\!\!\sous{1}{\hat
f}_0^{~j]L}+\!\!\!\sous{1}{f}_{-1}^{~j]L}\Bigr]\delta_1\biggr)+1\leftrightarrow
2\;.
\end{equation}
Remarkably, ${\sc H}_{[ij]}[F]$ takes exactly the same form as
(\ref{121'''}). Hence, we are able to determine a relation to be
satisfied by the looked-for coefficients $\alpha_l$ for any $l$ in
order that the non-commuting part (\ref{121'''}) associated to the
particular solution be cancelled out by that of the homogeneous one:
$\forall l$, $(l+2)\alpha_l-(l+1)\alpha_{l+1}=-2\pi(2l+3)$. Given any
initial value for $\alpha_0$ the solution reads as

\begin{equation}\label{122'}
\alpha_l=(l+1)\Biggl[\alpha_0+2\pi\sum_{j=1}^l\biggl({1\over j}+{1\over j+1}\biggr)\Biggr]
=-2\pi+4\pi(l+1)\biggl[K+\sum_{j=1}^l{1\over j+1}\biggr]\;,
\end{equation}
in which we have introduced the new arbitrary constant
$K={\alpha_0\over 4\pi}+{1\over 2}$. Inserting (\ref{122'}) back into
the expression for ${\sc D}_i[F]$ leads to the announced result
(\ref{121}). At last, we find that for any choice of the constant $K$
the second-derivative operator commutes, i.e.

\begin{equation}\label{122''}
{\sc D}_{[ij]}[F] ={\sc H}_{[ij]}[F]+{\sc D}_{[ij]}^{\rm part}[F]=0\;.
\end{equation}
Let us verify from (\ref{122''}) that all higher-order
multi-derivative operators commute as well, i.e. ${\sc
D}_{i_1i_2\cdots i_l}[F]$ given by the formula (\ref{120'}) is
symmetric in all its indices. This is easily proved by induction over
$l$. Suppose that to the $(l-1)$th order ${\sc D}_{i_1i_2\cdots
i_{l-1}}[F]$ is symmetric, and re-write the formula (\ref{120'}) into
both forms

\begin{eqnarray*} 
{\sc D}_{i_1i_2\cdots i_l}[F]&=&{\sc D}_{i_1}[\partial_{i_2\cdots
i_l}F]+\partial_{i_1}{\sc D}_{i_2\cdots i_l}[F]\\ &=&{\sc
D}_{i_1\cdots i_{l-1}}[\partial_{i_l}F]+\partial_{i_1\cdots
i_{l-1}}{\sc D}_{i_l}[F]\;.
\end{eqnarray*}
Clearly, ${\sc D}_{i_1\cdots i_l}[F]$ is symmetric with respect to
both $i_1\cdots i_{l-1}$ and $i_2\cdots i_l$, so must be symmetric in
all its indices (the symmetry with respect to the first and last
indices being a consequence of the other symmetries). {\it QED}.

We should mention that the dependence upon the arbitrary constant $K$
of the derivative operator defined by Theorem 4 is

\begin{equation}\label{123}
{\sc D}_i[F]_{~\bigl|_K}=4\pi K ~\!\sum_{l=0}^{+\infty}(l+1)~\!{\rm
Pf}~\!\biggl(\Bigl[n_1^{iL}\!\!\!\sous{1}{{\hat
f}}_{-1}^{~L}-n_1^L\!\!\!\sous{1}{{\hat
f}}_{-1}^{~iL}\Bigr]r_1\delta_1\biggr)+1\leftrightarrow 2\;,
\end{equation}
which can also be cast into the more interesting form

\begin{equation}\label{123'}
{\sc D}_i[F]_{~\bigl|_K}=-4\pi K ~\!\partial_i \Bigl[{\rm
Pf}(r_1^2\!\!\!\sous{1}{f}_{-1}\delta_1)\Bigr]+1\leftrightarrow 2\;.
\end{equation}
We see that the ``ambiguity'' linked with the constant $K$ when
deriving the pseudo-function ${\rm Pf}F$ is related to an ambiguity
resulting from the addition of the term $-4\pi K~\! {\rm
Pf}(r_1^2~\!{}_1f_{-1}\delta_1)+1\leftrightarrow 2$ to ${\rm Pf}F$. In
a sense, one can also view the constant $K$ as a measure of how much
the distributional derivative of the pseudo-function ${\rm Pf}(1/r_1)$
differs from the ordinary one, i.e.

\begin{equation}\label{124}
{\sc D}_i\left[{1\over r_1}\right]=4\pi K~\!{\rm
Pf}\left(n_1^ir_1\delta_1\right)\;.
\end{equation}
Indeed, for functions which are more singular than a simple $1/r_1$,
there is no dependence on the constant $K$; see
e.g. (\ref{97})-(\ref{98}).

\subsection{The Laplacian operator}

Let us compute the second-derivative of ${\rm Pf}(1/r_1)$ using the
formula ${\sc D}_{ij}[1/r_1]={\sc D}_i[-n_1^j/r_1^2]+\partial_i{\sc
D}_j[1/r_1]$. The first term is obtained directly from the definition
(\ref{121}), and the second term is computed with the help of the
formula (\ref{111}) applied on (\ref{124}).  As a result, we get
 
\begin{equation}\label{124'}
{\sc D}_{ij}\left[{1\over r_1}\right]=-{4\pi\over 3}~\!{\rm
Pf}\biggl(\left[\delta^{ij}+3(3K+1){\hat
n}_1^{ij}\right]\delta_1\biggr)\;,
\end{equation}
where ${\hat n}_1^{ij}=n_1^in_1^j-{1\over 3}\delta^{ij}$. Evidently
(because of the trace-free ${\hat n}_1^{ij}$), when we restrict
ourselves to smooth functions of the set ${\cal D}$, we recover the
usual formula of distributional theory,

\begin{equation}\label{124''}
{\sc D}_{ij}\left[{1\over r_1}\right]_{\!~\bigl|_{\cal D}}
=-{4\pi\over 3}~\!\delta^{ij}~\!\delta_1\;.
\end{equation}
Since the dependence over $K$ in (\ref{124'}) drops out when taking
the trace over the indices $ij$, we have ${\sc
D}_{ii}[1/r_1]=-4\pi{\rm Pf}\delta_1$ (even on the set ${\cal
F}$). This means that the Laplacian of $1/r_1$ on ${\cal F}$ takes the
same form as the well-known formula of distribution theory:

\begin{equation}\label{125}
\Delta\left({\rm Pf}{1\over r_1}\right)=-4\pi~\!{\rm Pf}\delta_1\;.
\end{equation}
We infer from the rule of integration by parts that

\begin{equation}\label{125'}
<\Delta({\rm Pf}F),{1\over r_1}>=<\Delta\left({\rm Pf}{1\over
r_1}\right),F>=-4\pi~\!(F)_1\;,
\end{equation}
which can be phrased by saying that the Poisson integral of the
Laplacian of a singular function, as evaluated at a singular point, is
equal to the partie finie of the function at that point.  More
generally, the Laplacian acting on any pseudo-function in ${\cal F}'$
is defined by

\begin{equation}\label{126}
\Delta({\rm Pf}F)={\rm Pf}(\Delta F)+{\sc D}_{ii}[F]\;,
\end{equation}
where the distributional term is given by

\begin{equation}\label{126'}
{\sc D}_{ii}[F]=\partial_i{\sc D}_i[F]+{\sc D}_i[\partial_iF]\;.
\end{equation}

\begin{proposition}
Under the hypothesis of Theorem 4 the distributional term associated
with the Laplacian operator reads

\begin{equation}\label{127}
{\sc D}_{ii}[F]=4\pi\sum_{l=0}^{+\infty}~\!{\rm
Pf}\Biggl((l+1)C_{l-1}~\!n_1^L\biggl[\!\!\sous{1}{{\hat
f}}_{-1}^{~L}+r_1\!\!\!\sous{1}{{\hat f}}_0^{~L}\biggr] \delta_1
-\sum_{k\geq 0}(2k+1){n_1^L\over r_1^k}\!\!\!\sous{1}{{\hat
f}}_{-1-k}^{~L}\delta_1\Biggr)+1\leftrightarrow 2\;.
\end{equation}
\end{proposition}
The proof is straightforward and will not be detailed. Note that the
dependence on $K$ occurs only for functions owing some non-zero
coefficients ${}_1f_{-1}$ or ${}_1f_0$, or $1\leftrightarrow 2$; for
instance

\begin{eqnarray*}
{\sc D}_{ii}[n_1^j]&=&8\pi K~\!{\rm
Pf}\left(n_1^ir_1\delta_1\right)\;,\\ {\sc D}_{ii}[n_1^j/r_1]&=&8\pi
(K-\case{1}{2})~\!{\rm Pf}\left(n_1^i\delta_1\right)\;.
\end{eqnarray*}
But, for more singular functions like $1/r_1^3$, we have

\begin{equation}\label{128}
\Delta\left({\rm Pf}{1\over r_1^3}\right)={\rm Pf}\left({6\over r_1^5}
-{20\pi\over r_1^2}\delta_1\right)\;.
\end{equation}

\begin{lemma}The Laplacian of the pseudo-function ${\rm Pf}(F\delta_1)$ 
is given by  

\begin{equation}\label{138}
\Delta\Bigl[{\rm Pf}(F\delta_1)\Bigr]=
{\rm Pf}\Biggl(r_1^3\Delta\biggl[{F\over
r_1^3}\biggr]\delta_1\Biggr)\;.
\end{equation}
\end{lemma}
The proof is similar to the one of the formula (\ref{111}).  Two
immediate particular applications are

\begin{mathletters}\label{139}\begin{eqnarray}
&&\Delta\Bigl({\rm Pf}\delta_1\Bigr)= {\rm Pf}\biggl({6\over
r_1^2}\delta_1\biggr)\;,\label{139a}\\ &&\Delta\Bigl[{\rm
Pf}(r_1^2\delta_1)\Bigr]=0\;,\label{139b}
\end{eqnarray}\end{mathletters}$\!\!$
which can also be deduced respectively from (\ref{20}) and
(\ref{18}). [(\ref{139a}) is in agreement with (\ref{77}).] Let us add
that

\begin{equation}\label{140}
\Delta\Biggl[{\rm Pf}\biggl({r_1\over 2}\delta_1\biggr)\Biggr]
={\rm Pf}\Biggl({\delta_1\over r_1}\Biggr) \;.
\end{equation}
In practice, Lemma 6 may be used to determine some solutions of
Poisson equations ``in the sense of distributions'' on ${\cal F}$. For
instance combining (\ref{139a}) with the formula (\ref{128}), we can
write

\begin{equation}\label{141}
\Delta\Biggl[{\rm Pf}\biggl({1\over r_1^3}+{10\pi\over 3}
\delta_1\biggr)\Biggr]={\rm Pf}\left({6\over r_1^5}\right)\;,
\end{equation}
which provides a solution of the Poisson equation with source ${\rm
Pf}(6/r_1^5)$ in the sense of these distributions. Such a solution is
by no means unique, since, from Lemma 6, one can add to it any
``homogeneous'' solution of the form ${\rm Pf}(H^{\rm hom}\delta_1)$
where $H^{\rm hom}$ is the product of $r_1^3$ with an arbitrary
solution of the Laplace equation. Notice that (\ref{141}) as it stands
is well-defined in distribution theory and so takes the same form when
restricted to ${\cal D}$ ($\Delta\delta_1$ is meaningful on this
set). However,

\begin{equation}\label{142}
\Delta\Biggl[{\rm Pf}\biggl({1\over r_1^2}+6\pi ~\!r_1\delta_1
\biggr)\Biggr]={\rm Pf}\left({2\over r_1^4}\right)
\end{equation}
has no equivalent in distribution theory.

\section{Time derivative and partial derivatives}

The functions $F\in {\cal F}$ depend on the field point ${\bf x}$ and
on the two singular source points ${\bf y}_1$ and ${\bf y}_2$. We
shall now consider the situation where the two source points represent
the trajectories of actual particles, and therefore depend on time
$t$. We assume that the two trajectories ${\bf y}_1(t)$ and ${\bf
y}_2(t)$ are smooth, that is ${\bf y}_1, {\bf y}_2 \in C^\infty
({\mathbb R})$. In general (e.g. in the application to the problem of
motion of point-particles) the function $F$ will also depend on time
through the two velocities ${\bf v}_1(t)=d{\bf y}_1(t)/dt$ and ${\bf
v}_2(t)=d{\bf y}_2(t)/dt$. We suppose that $F$ is a smooth functional
of ${\bf v}_1$ and ${\bf v}_2$. Therefore, in this section, $F$ is
supposed to take the form

\begin{equation}\label{143}
F=F\Bigl({\bf x},t;{\bf y}_1(t),{\bf y}_2(t)\Bigr)~\in~{\cal F}\;.
\end{equation}
We want to investigate the partial derivatives (in a distributional
sense) of the pseudo-function ${\rm Pf}F$ with respect to the source
points ${\bf y}_1$ and ${\bf y}_2$, as well as the derivative of ${\rm
Pf}F$ with respect to time $t$. Obviously, the partial derivatives
${}_1\partial_i\equiv\partial/\partial {\bf y}_1$ and
$1\leftrightarrow 2$ are closely related to the time derivative
$\partial_t\equiv\partial/\partial t$ on account of the fact that

\begin{equation}\label{144}
\partial_tF={\dot F}+v_1^i\!\!\sous{1}{\partial}_{~i}F+v_2^i
\!\!\sous{2}{\partial}_{~i}F\;,
\end{equation}
(in the ordinary sense), where ${\dot F}$ denotes the contribution of
the time-derivative due to the dependence over the velocities,
i.e. ${\dot F}=a_1^i~\!\partial F/\partial v_1^i+a_2^i~\!\partial
F/\partial v_2^i$ ($a_1^i$ and $a_2^i$ denoting the two
accelerations). In applications it is frequent that $F$ depends on the
trajectories only through the two distances to the field point ${\bf
r}_1={\bf x}-{\bf y}_1$ and ${\bf r}_2={\bf x}-{\bf y}_2$; in that
case

\begin{equation}\label{145}
\partial_iF+\!\!\sous{1}{\partial}_{~i}F+\!\!\sous{2}{\partial}_{~i}F=0\;.
\end{equation}
The general function (\ref{143}) does not necessarily satisfy the
latter identity. However, let us guess from (\ref{145}) the result for
the distributional terms ${}_1{\sc D}_i[F]$ (and $1\leftrightarrow 2$)
associated with the partial derivative ${}_1\partial_i$ acting on the
pseudo-function ${\rm Pf}F$. Since we have supposed that the
dependence of $F$ on the velocities is smooth, the distributional
terms will depend only on that part of the function which becomes
singular when $r_1\to 0$, and so, because as far as the singular part
is concerned, the function behaves like (\ref{145}), the
distributional terms ${}_1{\sc D}_i[F]$ and ${}_2{\sc D}_i[F]$ should
satisfy

\begin{equation}\label{150}
{\sc D}_i[F]+\!\!\sous{1}{{\sc D}}_{~i}[F]+\!\!\sous{2}{{\sc D}}_{~i}[F]=0\;.
\end{equation}
Now, from Theorem 4, we know that ${\sc D}_i[F]$ can be naturally
split into two parts associated respectively with the singularities 1
and 2. Therefore, we expect that the correct distributional term
${}_1{\sc D}_i[F]$ is equal to {\it minus} that part of ${\sc D}_i[F]$
which corresponds to 1. Namely, using (\ref{121}),

\begin{equation}\label{150'}
\sous{1}{{\sc D}}_{~i}[F]=-4\pi\sum_{l=0}^{+\infty}~\!{\rm Pf}
\Biggl(~\!C_l~\!\biggl[n_1^{iL}\!\!\!\sous{1}{{\hat f}}_{-1}^
{~L}-n_1^{L}\!\!\!\sous{1}{{\hat f}}_{-1}^{~iL}\biggr] r_1\delta_1
+\sum_{k\geq 0}{n_1^{iL}\over r_1^k}\!\!\!\sous{1}{{\hat
f}}_{-2-k}^{~L}\delta_1\Biggr)
\end{equation}
(and {\it idem} for 2). This expectation is confirmed by the following
definition and proposition.

\begin{definition} 
The partial derivative $\!\!\sous{\!\!1}{~\partial}_{\!~~i}$ (and
$1\leftrightarrow 2$) acting on pseudo-functions is said to satisfy
the rule of integration by parts iff

\begin{equation}\label{146}
\forall F,G\in{\cal F}\;,\quad
<\!\!\sous{1}{\partial}_{~i}({\rm Pf}F),G>+
<\!\!\sous{1}{\partial}_{~i}({\rm
Pf}G),F>=\!\!\sous{1}{\partial}_{~i}\Bigl[<{\rm Pf}F,G>\Bigr]\;.
\end{equation}
Similarly, the time derivative $\partial_t$ is said to satisfy the
rule of integration by parts iff

\begin{equation}\label{147}
<\partial_t({\rm Pf}F),G>+ <\partial_t({\rm Pf}G),F>={d\over
dt}\Bigl[<{\rm Pf}F,G>\Bigr]\;.
\end{equation}
\end{definition}
Notice that $<{\rm Pf}F,G>={\rm Pf}\int d^3{\bf x}~\!FG$ is a function
of the source points ${\bf y}_1(t)$ and ${\bf y}_2(t)$, as well as $t$
independently if either $F$ or $G$ depends on the velocities. The time
derivative in the right side of (\ref{147}) means the total time
derivative we get by taking into account both the variable $t$
occuring through ${\bf y}_1(t)$ and ${\bf y}_1(t)$, and the
independent $t$ coming from the velocities. Let us now state a result
analogous to Theorem 4, whose proof will not be given since it
represents a simple adaptation of the one of that theorem.

\begin{proposition} 
Under the hypothesis of Theorem 4 the partial derivative with respect
to ${\bf y}_1$ (and idem with $1\leftrightarrow 2$) is determined as

\begin{equation}\label{148}
\sous{1}{\partial}_{~i}({\rm Pf}F)={\rm Pf}(\!\!\sous{1}{\partial}
_{~i}F)+\!\!\sous{1}{{\sc D}}_{~i}[F]\;,
\end{equation}
where ${}_1{\sc D}_i[F]$ is given by ${\rm (\ref{150'})}$.  And the
time derivative is determined as

\begin{equation}\label{149a}
\partial_t({\rm Pf}F)={\rm Pf}(\partial_tF)+{\sc D}_t[F]\;,
\end{equation}
where ${\sc D}_t[F]$ is given by

\begin{equation}\label{149b}
{\sc D}_t[F]=v_1^i\!\!\sous{1}{{\sc D}}_{~i}[F]+v_2^i\!\!\sous{2}{{\sc
D}}_{~i}[F]\;.
\end{equation}
\end{proposition}
Higher-order derivatives are constructed as in Section VIII. We find
for instance

\begin{equation}\label{151}
\sous{1}{\partial}_{~ij}({\rm Pf}F)={\rm Pf}(\!\!\sous{1}{\partial}_
{~ij}F)+\!\!\sous{1}{{\sc D}}_{~i}[\!\!\sous{1}{\partial}_{~j}F]
+\!\!\sous{1}{\partial}_{~i}\!\!\!\sous{1}{{\sc D}}_{~j}[F]\;,
\end{equation}
{\it Idem} for the second-order time derivative, which reads as

\begin{equation}\label{152}
\partial_t^{\!\!~2}({\rm Pf}F)={\rm Pf}(\partial_t^{\!\!~2}F)
+{\sc D}_t[\partial_tF]+\partial_t{\sc D}_t[F]\;,
\end{equation}
where $\partial_tF$ is given by (\ref{144}) and ${\sc D}_t[F]$ is
defined in (\ref{149b}). Furthermore the mixing up of derivatives of
different type is allowed, and proceeds in the expected way. E.g.

\begin{equation}\label{153}
\sous{1}{\partial}_{\!~i}\!\!\!\!\sous{2}{\partial}_{\!~j}
({\rm Pf}F)={\rm Pf}(\!\!\sous{1}{\partial}_{\!~i}\!\!\!\!\sous{2}
{\partial}_{\!~j}F)
+\!\!\sous{1}{{\sc D}}_{\!~i}[\!\!\sous{2}{\partial}_{\!~j}F]
+\!\!\sous{1}{\partial}_{\!~i}\!\!\!\!\sous{2}{{\sc D}}_{\!~j}[F]\;.
\end{equation}
Another example is

\begin{equation}\label{154}
\partial_t\partial_{ij}({\rm Pf}F)={\rm Pf}(\partial_t\partial_{ij}F)
+{\sc D}_t[\partial_{ij}F]+\partial_t{\sc
D}_i[\partial_jF]+\partial_t\partial_i{\sc D}_j[F]\;.
\end{equation}

\acknowledgments

The authors are grateful to Antoine Sellier for discussion and his
interesting comments. This work was supported in part by the National
Science Foundation under Grant no PHY-9900776.

\appendix

\section{Proof of Theorem 2}

Basically the proof establishes the legitimacy of commuting some
discrete series with integrals.  Consider $F\in{\cal F}$. We start by
evaluating the integrals

\begin{mathletters}\label{155}\begin{eqnarray}
&& {\rm Pf}_{\beta \to 0 \atop \alpha \to 0}
\int_{{\cal B}_1(s)}  
d^3{\bf x}~\sum_{a \le -3}
\left(\frac{r_1}{s_1}\right)^\alpha
\left(\frac{r_2}{s_2} \right)^\beta r_1^a \! \! \! \!
\sous{1}{f}_a({\bf n}_1) \label{155a}\\  
\mathrm{and} \qquad && 
{\rm Pf}_{\alpha \to 0 \atop \beta \to 0}
\int_{{\cal B}_1(s)} d^3{\bf x}~\sum_{a \le -3}
\left(\frac{r_1}{s_1}\right)^\alpha
\left(\frac{r_2}{s_2} \right)^\beta r_1^a \! \! \! \!
\sous{1}{f}_a({\bf n}_1)\;, \label{155b}
\end{eqnarray}\end{mathletters}$\!\!$
where the ${}_1f_a$'s are the coefficients of the expansion of $F$
when $r_1\to 0$, and where ${\cal B}_1(s)$ is the ball centred on
${\bf y}_1$ and of radius $s\in {\mathbb R}^{+*}$ (chosen to be
$s<r_{12}$).  From the definition of the class ${\cal F}$ the sums
over $a$ in (\ref{155}) are finite. When the real part of $\beta$ is
such that $0 \le \Re{(\beta)} \le 1$, the integrand of (\ref{155a}) is
majored by

$$\left(\frac{r_1}{s_1}\right)^{\Re(\alpha)}
\left(\frac{r_2}{s_2}\right)^{\Re(\beta)} \sum_{a \le -3} r_1^a |\! \! \! \!
\sous{1}{f}_a({\bf n}_1) |
\le \left(\frac{r_1}{s_1}\right)^{\Re(\alpha)}
\max\left(1,\frac{r_2}{s_2}\right) \sum_{a \le -3} r_1^a |\! \! \! \!
\sous{1}{f}_a({\bf n}_1) |\; , $$ 
which can be integrated on ${\cal B}_1(s)$. Thus the theorem of
dominated convergence of an integral can be applied, with the result
that

\begin{eqnarray*} 
&& {\rm Pf}_{\beta \to 0 \atop \alpha \to 0}
\int_{{\cal B}_1(s)} \! \! d^3{\bf x}~\sum_{a \le -3}
\left(\frac{r_1}{s_1}\right)^\alpha
\left(\frac{r_2}{s_2} \right)^\beta r_1^a \! \! \! \!
\sous{1}{f}_a = {\rm Pf}_{\alpha \to 0}
\int_{{\cal B}_1(s)} \! \! d^3{\bf x}~\sum_{a \le -3}  
\left(\frac{r_1}{s_1}\right)^\alpha r_1^a \! \! \! \!
\sous{1}{f}_a \\ & & 
\qquad\qquad\qquad =  \sum_{a+3 < 0} \frac{s^{a+3}}{a+3} \int d\Omega_1 
\! \! \! \! \sous{1}{f}_a +
\ln \left(\frac{s}{s_1} \right) \int d\Omega_1 \! \! \! \!
\sous{1}{f}_{-3} \; . 
\end{eqnarray*}
The second integral is more difficult to compute because the limit
$\alpha \to 0$ does not commute with the integration sign. We must
expand $r_2^\beta$ as a power series of $r_1$. $\forall r_1 < r_{12}$,

\begin{equation}\label{156}
r_2^\beta=r_{12}^\beta \biggl(1+2 {\bf n}_1.{\bf n}_{12}
\frac{r_1}{r_{12}}+
\frac{r_1^2}{r_{12}^2}\biggr)^{\beta/2}
=r_{12}^\beta \sum_{l=0}^{+\infty} C^{-\beta/2}_l\!\left(-{\bf n}_1.
{\bf n}_{12}\right) \left( \frac{r_1}{r_{12}} \right)^l \; ,
\end{equation}
where ${\bf n}_{12}=({\bf y}_1-{\bf y}_2)/r_{12}$, and where
$C^\lambda_l(t)$ denotes the Gegenbauer polynomial, which is by
definition the coefficient of $x^l$ in the power-series expansion of
the function $(1-2 t x+x^2)^{-\lambda}$ when $x\to 0$ (with $\lambda$,
$t\in {\mathbb C}$). See e.g. Morse and Feshbach\cite{MorseF}
p. 602. When $t\in {\mathbb R}$ and is such that $|t|\leq 1$ (as is
the case here since $t=-{\bf n}_1.{\bf n}_{12}$), we can obtain a
majoration of the Gegenbauer polynomial. From the formula ({\it cf}
Gradshteyn and Ryzhik \cite{GR} p. 1030)

$$C^\lambda_l(\cos \theta)=\sum_{k,h \ge 0 \atop k+h=l}
\frac{\Gamma(\lambda+k) \Gamma(\lambda+h)}{k!~\! h!~\! [\Gamma(\lambda)]^2} 
\cos [(k-h)\theta]\;,
$$ we find that $\forall l \neq 0$, $|C^\lambda_l(\cos \theta)|$ is
always less than
 
$$\sum_{k,h \ge 0 \atop k+h=l} \frac{(|\lambda|+k-1)(|\lambda|+k-2)
\cdots |\lambda| (|\lambda|+h-1)(|\lambda|+h-2) \cdots |\lambda|}{k!~\!
h!}=C^{|\lambda|}_l(1) \; .$$ Therefore the series $\sum_l |
C^{-\beta/2}_l(-{\bf n}_1.{\bf n}_{12})) (r_1/r_{12})^l|$ is bounded
by $(1-2r_1/r_{12}+r_1^2/r_{12}^2)^{|\beta|/2}$, and thus admits a
limit. Thus (\ref{156}) converges absolutly and (when $\Re (\alpha)$
is large enough) the signs $\int$ and $\sum$ can be interchanged:
 
$$ 
\int_{{\cal B}_1(s)} d^3{\bf x}~\sum_{a+3 \le 0}
\left(\frac{r_1}{s_1}\right)^\alpha
\left(\frac{r_2}{s_2} \right)^\beta r_1^a \! \! \!
\sous{1}{f}_a
=\left(\frac{r_{12}}{s_2} \right)^\beta
\sum_{l=0}^{+\infty} \sum_{a+3 \le 0} \int_{{\cal B}_1(s)}
d^3{\bf x}~\frac{r_1^{\alpha+a+l}}{s_1^\alpha r_{12}^l}
\! \! \! \sous{1}{f}_a
~C^{-\beta/2}_l\;, $$ where $C^{-\beta/2}_l\equiv C^{-\beta/2}_l(-{\bf
n}_1.{\bf n}_{12})$.  We obtain the two terms

\begin{eqnarray*}
&&\left(\frac{r_{12}}{s_2} \right)^\beta \sum_{l=0}^{+\infty}
\sum_{a+3 \le 0 \atop a+l+3 \neq 0}
\frac{s^{\alpha+a+l+3}}{s_1^\alpha 
r_{12}^l (\alpha+a+l+3)} \int d\Omega_1 \! \! \! \sous{1}{f}_a
~C^{-\beta/2}_l \\
&&\qquad\qquad\qquad\qquad\qquad +
\left(\frac{r_{12}}{s_2} \right)^\beta \sum_{l=0 \atop
{\rm finite~sum}}^{+\infty} 
\frac{1}{\alpha} \left(\frac{s}{s_1} \right)^\alpha
\frac{1}{r_{12}^l} \int d\Omega_1 \! \! \! \sous{1}{f}_{-l-3}
~C^{-\beta/2}_l \; .
\end{eqnarray*}
The finite part when $\alpha\to 0$ of the second term reads simply

\begin{equation}\label{157} \left(\frac{r_{12}}{s_2} \right)^\beta \ln \left(
\frac{s}{s_1} \right) 
\sum_{l=0}^{+\infty} \frac{1}{r_{12}^l} \int d\Omega_1
\! \! \! \sous{1}{f}_{-l-3}~C^{-\beta/2}_l\; . \end{equation} 
On the other hand, in order to treat the first term, we must justify
the commutation of the finite part with the infinite sum. Consider the
series

$$\sum_{l=0}^{+\infty} \frac{1}{\alpha+a+l+3}
\left(\frac{s}{r_{12}}\right)^l \int d\Omega_1
\! \! \!
\sous{1}{f}_a~C^{-\beta/2}_l\; .
$$ For $\alpha$ in a disk of the complex plane centered on $0$ and of
radius $\epsilon$ (with $0<\epsilon<1$), we can bound the generic term
of that series (for large enough $l$) by

$$\frac{1}{|a+l+3|-\epsilon} ~\left(\frac{s}{r_{12}}\right)^l
\bigg|\int d\Omega_1
\! \! \!
\sous{1}{f}_a~C^{-\beta/2}_l\bigg| \;,
$$ which is independent of $\alpha$, and whose corresponding series in
$l$ converges.  Therefore we can apply the limit $\alpha\to 0$ through
the summation over $l$ and deduce

\begin{eqnarray}\label{158}
&& \lim_{\alpha \to 0} \Biggl\{ \left(\frac{r_{12}}{s_2} \right)^\beta
\sum_{l=0}^{+\infty}
\sum_{a+3 \le 0 \atop a+l+3 \neq 0}
\frac{s^{\alpha+a+l+3}}{s_1^\alpha 
r_{12}^l (\alpha+a+l+3)} \int d\Omega_1 \! \! \! \sous{1}{f}_a
~C^{-\beta/2}_l\Biggr\} \nonumber\\ 
&&\qquad\qquad\qquad =\left(\frac{r_{12}}{s_2} \right)^\beta \sum_{a+3 \le 0}
s^{a+3} \sum_{l=0 \atop a+l+3 \neq 0}^{+\infty} 
\frac{1}{a+l+3} \left(\frac{s}{r_{12}}\right)^l 
\int d\Omega_1 \! \! \! \sous{1}{f}_a
~C^{-\beta/2}_l \;.
\end{eqnarray}
Next we apply the finite part ${\rm Pf}_{\beta \to 0}$ to the sum of
(\ref{157}) and (\ref{158}), which involves finding the limit when
$\beta \to 0$ of the series

\begin{equation}\label{159} 
\sum_{l=0 \atop a+l+3 \neq 0}^{+\infty} 
\frac{1}{a+l+3} \left(\frac{s}{r_{12}}\right)^l 
\int d\Omega_1 \! \! \! \sous{1}{f}_a
~C^{-\beta/2}_l \;.
\end{equation}
In any case the absolute value of the quantity under the sign $\sum$
is smaller than

$$\left(\frac{s}{r_{12}}\right)^l 
C^{|\beta|/2}_l(1) 
\int d\Omega_1 ~\!|\!\! \! \! \sous{1}{f}_a| \;.
$$ Furthermore we know that $C^\lambda_l(1)=\Gamma (2\lambda+l)/[l!~\!
\Gamma (2\lambda)]$.  For $l \not= 0$, $C^{|\lambda|}_l(1)=(2
|\lambda|+l-1) (2 |\lambda|+l-2) \cdots (2 |\lambda|)/~\! l!$ is
manifestly an increasing function of $|\lambda|$, and, for $l=0$,
$C^{|\lambda|}_0(1)=1$ is constant. From this we infer that $\forall
l$ and for $\beta$ in the disk centred on $0$ and of radius
$\epsilon$, $C^{|\beta|/2}_l(1) \le C^{\epsilon/2}_l(1)$ holds, which
leads to the $\beta$-independent bound

$$\left(\frac{s}{r_{12}}\right)^l C^{\epsilon/2}_l(1)
\int d\Omega_1~\! |\!\! \! \! \sous{1}{f}_a| \;,
$$ which is manifestly the general term of a convergent series.
Therefore the series (\ref{159}) possesses a limit when $\beta\to 0$
which is simply obtained by setting $\beta=0$ under the sign
$\sum$. Using $C^0_l(\cos \theta)=\delta_{l0}$ we find this limit to
be $0$ if $a=-3$ and
 
\begin{equation}\label{160}  
\frac{1}{a+3} \int d\Omega_1 \! \! \! 
\sous{1}{f}_a
\end{equation} 
if $a \not= -3$.  Gathering the results (\ref{157})-(\ref{158}) and
(\ref{160}), we arrive to

\begin{eqnarray}\label{161} 
&& \qquad \sum_{a+3 < 0}
\frac{s^{a+3}}{a+3} \int d\Omega_1  
\! \! \! \! \sous{1}{f}_a+
\ln \left(\frac{s}{s_1} \right) \int d\Omega_1 \! \! \! \!
\sous{1}{f}_{-3}\qquad \qquad \qquad \qquad \qquad \quad
\mbox{} \nonumber \\ 
&& \qquad \qquad \qquad \qquad \qquad = {\rm FP}_{\alpha \to 0 \atop
\beta \to 0}
\int_{{\cal B}_1(s)}  d^3{\bf x} \sum_{a+3 \le 0}
\left(\frac{r_1}{s_1}\right)^\alpha 
\left(\frac{r_2}{s_2} \right)^\beta  r_1^a \! \! \! \!
\sous{1}{f}_a \nonumber \\ 
&& \qquad \qquad \qquad \qquad \qquad = {\rm FP}_{\beta \to 0 \atop
\alpha \to 0}
\int_{{\cal B}_1(s)} d^3{\bf x} \sum_{a+3 \le 0}
\left(\frac{r_1}{s_1}\right)^\alpha 
\left(\frac{r_2}{s_2} \right)^\beta  r_1^a \! \! \! \!
\sous{1}{f}_a\;, 
\end{eqnarray}  
from which we can now easily prove the equivalence with the Hadamard
partie finie. Like in the proof of Proposition 1 we consider two open
domains ${\cal D}_1$ and ${\cal D}_2$, disjoined and complementary in
${\mathbb R}^3$, and such that ${\cal B}_1(s)\subset {\cal D}_1$ and
${\cal B}_2(s)\subset {\cal D}_2$. We can write

\begin{equation}\label{162}
\int_{{\cal D}_1} d^3{\bf x}~\left(\frac{r_1}{s_1}
\right)^\alpha \! \! \left(\frac{r_2}{s_2} \right)^\beta F=
\int_{{\cal D}_1\setminus {\cal B}_1(s)} \! \! \!
d^3{\bf x}~\left(\frac{r_1}{s_1}
\right)^\alpha \left(\frac{r_2}{s_2} \right)^\beta F+
\int_{{\cal B}_1(s)} \! \! \! 
d^3{\bf x}~\left(\frac{r_1}{s_1} 
\right)^\alpha \! \! \left(\frac{r_2}{s_2} \right)^\beta F\;,
\end{equation}
where each of the objects is defined by complex analytic continuation
in a neighbourhood of $\alpha=\beta=0$ [proof similar to the one after
(\ref{30})].  Like in (\ref{40}) we associate to $F$ the function
${\widetilde F}_1$ representing its ``regularization'' around the
point 1,

\begin{equation}\label{163}
{\widetilde F}_1 = F-\sum_{a+3 \leq 0} r_1^a
\! \! \sous{1}{f}_a\;,
\end{equation} 
and we re-write the right side of (\ref{162}) as

$$\int_{{\cal D}_1\setminus {\cal B}_1(s)} \! \! \!  d^3{\bf
x}~\left(\frac{r_1}{s_1}
\right)^\alpha \left(\frac{r_2}{s_2} \right)^\beta F
+\int_{{\cal B}_1(s)} \! \! \! 
d^3{\bf x}~\left(\frac{r_1}{s_1} 
\right)^\alpha \! \! \left(\frac{r_2}{s_2} \right)^\beta 
{\widetilde F}_1
+\int_{{\cal B}_1(s)} \! \! \!  d^3{\bf x}~\left(\frac{r_1}{s_1}
\right)^\alpha \! \! \left(\frac{r_2}{s_2} \right)^\beta 
\sum_{a+3 \leq 0} r_1^a  
\! \! \sous{1}{f}_a\;.
$$ Of these three terms, the first two are well-defined when $\alpha$
and $\beta$ tend to zero, hence their finite parts are simply obtained
by posing $\alpha=0=\beta$. On the other hand we have proved
previously that the finite parts ${\rm Pf}_{\alpha \to 0
\atop \beta \to 0}$ and ${\rm Pf}_{\beta \to 0
\atop \alpha \to 0}$ of the third term are equal and 
we have found their common value to be given by (\ref{161}). 
This shows immediately that 

\begin{eqnarray}\label{164}
{\rm FP}_{\alpha \to 0
\atop \beta \to 0} \int_{{\cal D}_1} d^3{\bf x}~\left(\frac{r_1}{s_1}
\right)^\alpha \! \! \left(\frac{r_2}{s_2} \right)^\beta F&=&
\int_{{\cal D}_1\setminus {\cal B}_1(s)} \! \! \!
d^3{\bf x}~F+\int_{{\cal B}_1(s)} \! \! \! 
d^3{\bf x}~{\widetilde F}_1\nonumber\\
&+&\sum_{a+3 < 0}
\frac{s^{a+3}}{a+3} \int d\Omega_1  
\! \! \! \! \sous{1}{f}_a+
\ln \left(\frac{s}{s_1} \right) \int d\Omega_1 \! \! \! \!
\sous{1}{f}_{-3}
\end{eqnarray}
(and {\it idem} with ${\rm Pf}_{\beta \to 0
\atop \alpha \to 0}$).
We recognize in the right side of (\ref{164}) the Hadamard partie
finie of the integral. Indeed the second term clearly admits an
expansion in positive powers of $s$,

\begin{equation}\label{165}
\forall N\in {\mathbb N}\;,\quad\int_{{\cal B}_1(s)} \! \! \! 
d^3{\bf x}~{\widetilde F}_1 = \sum_{0< a+3 \leq N}
\frac{s^{a+3}}{a+3} \int d\Omega_1  
\! \! \! \! \sous{1}{f}_a + o(s^N)\;,
\end{equation}
so we recover exactly the partie-finie integral over ${\cal D}_1$ in
the form given by (\ref{22'}). {\it QED}.

\end{document}